\documentclass{nature}

\newcounter{firstbib}

\usepackage{graphicx}
\usepackage{deluxetable}
\usepackage{epsfig}
\usepackage{tablefootnote}
\usepackage{url}
\usepackage{amsmath,times}
\usepackage{lineno}

\usepackage{caption}
\usepackage{subcaption}

\usepackage{amssymb}
\usepackage{amsmath}
\usepackage{tablefootnote}
\DeclareMathOperator{\arcsec}{arcsec}

\newcommand{\aap}{Astron. Astrophys.}
\newcommand{\araa}{Ann. Rev. Astron. Astrophys.}
\newcommand{\apj}{Astrophys. J.}
\newcommand{\aj}{Astron. J.}

\newcommand{\apjs}{Astrophys. J. Supplements}
\newcommand{\nat}{Nature}
\newcommand{\pasp}{Publications of the Astronomical Society of the Pacific}

\newcommand{\mnras}{Mon. Not. R. Astron. Soc.}

\title{A WC/WO star exploding within an expanding carbon-oxygen-neon nebula}

\author{A.~Gal-Yam,$^{1}$ 
 R.~Bruch,$^{1}$ 
 S.~Schulze,$^{1,2}$ 
 Y.~Yang,$^{1,3}$
 D.~A.~Perley,$^4$
 I.~Irani,$^1$
 J.~Sollerman,$^2$
 E.~C.~Kool,$^2$
 M.~T.~Soumagnac,$^{1,5}$
 O.~Yaron,$^1$
 N.~L.~Strotjohann,$^1$
 E.~Zimmerman,$^1$
 C.~Barbarino,$^2$
 S.~R.~Kulkarni,$^6$
 M.~M.~Kasliwal,$^6$
 K.~De,$^6$
 Y.~Yao,$^6$
 C.~Fremling,$^6$
 L.~Yan,$^6$
 E.~O.~Ofek,$^{1}$
 C.~Fransson,$^{2}$
 A.~V.~Filippenko,$^{3,7}$
 W.~Zheng,$^3$
 T.~G.~Brink,$^3$
 C.~M.~Copperwheat,$^4$
 R.~J.~Foley,$^8$
 J.~Brown,$^8$
 M.~Siebert,$^8$
 G.~Leloudas,$^9$
 A.~L.~Cabrera~Lavers,$^{10}$
 D.~Garcia~Alvarez,$^{10}$
 A.~Marante~Barreto,$^{10}$
 S.~Frederick,$^{11}$
 T.~Hung,$^8$
 J. C. Wheeler,$^{12}$
 J. Vink\'{o},$^{13,14,15,12}$
 B. P. Thomas,$^{12}$
 M.~J.~Graham,$^6$
 D.~A.~Duev,$^6$
 A.~J.~Drake,$^6$
 R.~Dekany,$^6$
 E.~C.~Bellm,$^{16}$
 B.~Rusholme,$^{17}$
 D.~L.~Shupe,$^{17}$
 I.~Andreoni,$^6$
 Y.~Sharma,$^6$
 R.~Riddle,$^6$
 J.~van~Roestel,$^6$
 N.~Knezevic,$^{18}$
 }

\begin{document}

\maketitle

\begin{affiliations}
   \item Department of Particle Physics and Astrophysics, Weizmann Institute of Science, 76100 Rehovot, Israel
   \item The Oskar Klein Centre, Department of Astronomy and Department of Physics, Stockholm University, AlbaNova, SE-106 91 Stockholm, Sweden
   \item Department of Astronomy, University of California, Berkeley, CA 94720-3411, USA
   \item Astrophysics Research Institute, Liverpool John Moores University, IC2, Liverpool Science Park, 146 Browlow Hill, Liverpool L3 5RF, UK
   \item Lawrence Berkeley National Laboratory, 1 Cyclotron Road, MS 50B-4206, Berkeley, CA 94720, USA
   \item Division of Physics, Mathematics, and Astronomy, California Institute of Technology, Pasadena, CA 91125, USA
   \item Miller Institute for Basic Research in Science, University of California, Berkeley, CA 94720, USA
   \item Department of \aap, University of California, Santa Cruz, CA 95064, USA
   \item DTU Space, National Space Institute, Technical University of Denmark, Elektrovej 327, DK-2800 Kgs. Lyngby, Denmark
   \item Grantecan S. A., Centro de Astrofísica de La Palma, Cuesta de San José, 38712 Breña Baja, La Palma, Spain
   \item Department of Astronomy, University of Maryland, College Park, MD 20742, USA
   \item Department of Astronomy, University of Texas at Austin, Austin, TX 78712, USA
   \item Konkoly Observatory, ELKH CSFK, Konkoly Thege M. ut 15-17, Budapest, 1121, Hungary
   \item Department of Optics \& Quantum Electronics, University of Szeged, Dom ter 9, Szeged, 6720 Hungary
   \item ELTE E\"otv\"os Lor\'and University, Institute of Physics, P\'azm\'any P\'eter s\'et\'any 1/A, Budapest, 1117, Hungary
   \item DIRAC Institute, Department of Astronomy, University of Washington, 3910 15th Avenue NE, Seattle, WA 98195, USA
   \item IPAC, California Institute of Technology, 1200 E. California
   Blvd, Pasadena, CA 91125, USA
   \item Department of Astronomy, Faculty of Mathematics, University of Belgrade, Studentski trg 16, 11000 Belgrade, Serbia
   
\end{affiliations}

%%%%%%%%%%%%%%%%%%%%%%%%%%%%%%%%%%%%%%%%%%%%%%%%%%%%%%%%%%%%%%%%%%%%%%%%%
%								ABSTRACT
%%%%%%%%%%%%%%%%%%%%%%%%%%%%%%%%%%%%%%%%%%%%%%%%%%%%%%%%%%%%%%%%%%%%%%%%%

\clearpage

\begin{abstract}

The final explosive fate of massive stars, and the nature of the compact remnants they leave behind (black holes and neutron stars), are major open questions in astrophysics. Many massive stars are stripped of their outer hydrogen envelopes as they evolve. Such Wolf-Rayet (W-R) stars\cite{Crowther2007} emit strong and rapidly expanding (v$_{\rm wind}>1000$\,km\,s$^{-1}$) winds indicating a high escape velocity from the stellar surface. A fraction of this population is also helium depleted, with spectra dominated by highly-ionized emission lines of carbon and oxygen (Types WC/WO). Evidence indicates that the most commonly-observed supernova (SN) explosions that lack hydrogen and helium (Types Ib/Ic) cannot result from massive WC/WO stars\cite{Smith2011a,Taddia2018}, leading some to suggest that most such stars collapse directly into black holes without a visible supernova explosion\cite{Dessart2011}. Here, we present observations of SN 2019hgp, discovered about a day after explosion. The short rise time and rapid decline place it among an emerging population of rapidly-evolving transients (RETs)\cite{Drout2014,Arcavi2016,Pursiainen2018,Perley2020}. Spectroscopy reveals a rich set of emission lines indicating that the explosion occurred within a nebula composed of carbon, oxygen, and neon. Narrow absorption features show that this material is expanding at relatively high velocities ($>1500$\,km\,s$^{-1}$) requiring a compact progenitor. Our observations are consistent with an explosion of a massive WC/WO star, and suggest that massive W-R stars may be the progenitors of some rapidly evolving transients.            
\end{abstract}

%%%%%%%%%%%%%%%%%%%%%%%%%%%%%%%%%%%%%%%%%%%%%%%%%%%%%%%%%%%%%%%%%%%%%%%%%
%								MAIN TEXT
%%%%%%%%%%%%%%%%%%%%%%%%%%%%%%%%%%%%%%%%%%%%%%%%%%%%%%%%%%%%%%%%%%%%%%%%%

\begin{figure}
 \centering
\begin{tabular}{cc}
\hspace*{-3cm}\includegraphics[width=21cm]{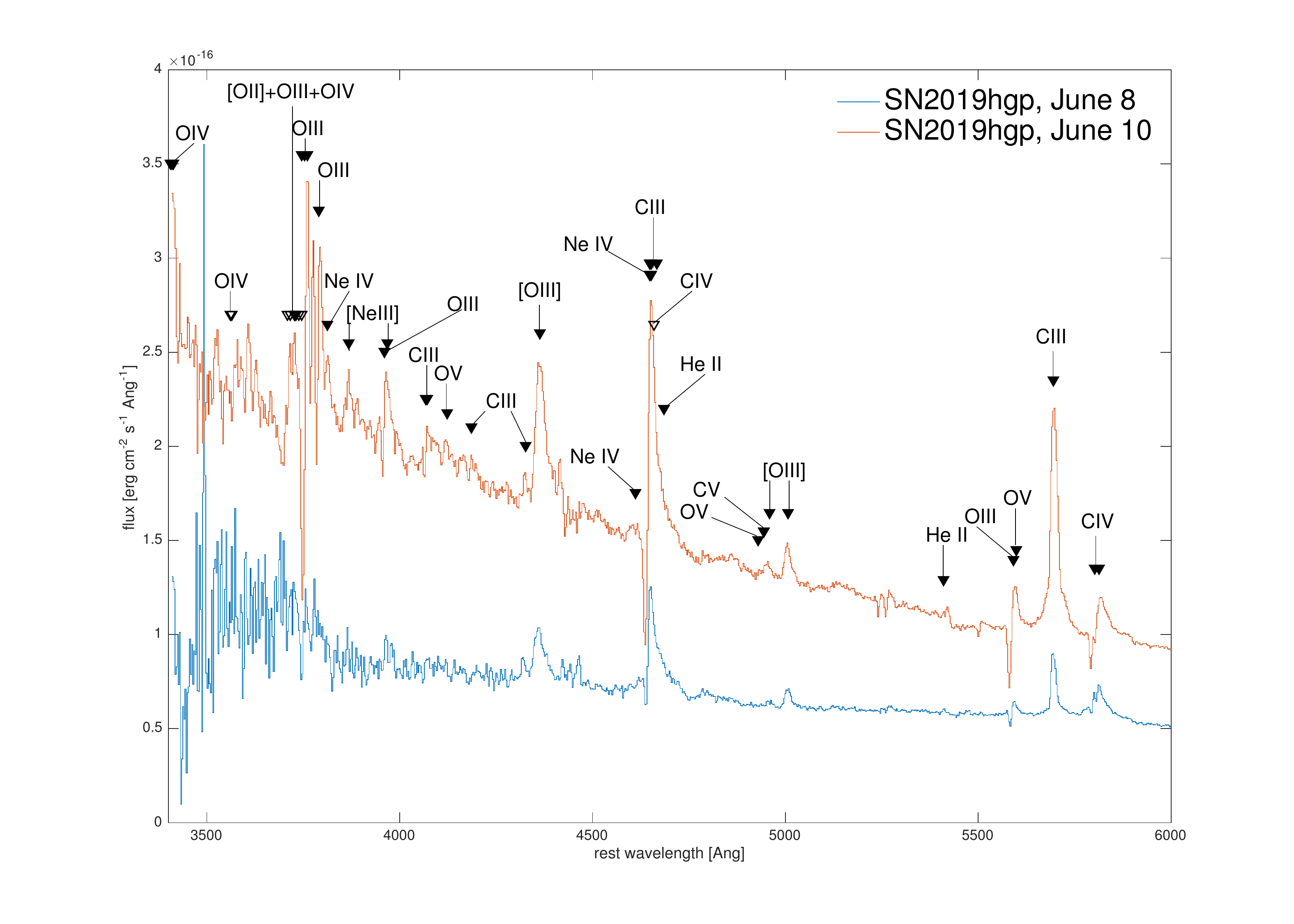}
\vspace*{-3cm}
\end{tabular}
\caption{
Spectra of SN 2019hgp are dominated by carbon, oxygen, and neon. High-quality spectra of SN 2019hgp obtained with Gemini/GMOS  only 1\,d and 3\,d after explosion are analyzed using the method of Gal-Yam\cite{agy18}, with all lines above $30\%$ of maximum intensity marked. The first spectrum is impacted by slit losses blueward of 4400\,\AA\ and its continuum was artificially made similar to that of the high-quality spectrum obtained 2\,d later. Almost all features are clearly associated with high-ionization transitions of C, O, and Ne. In particular, strong features of ionized He around 4686\,\AA\ and 5411\,\AA\ are not seen.
  \label{fig:spec}}
 \end{figure}

\begin{figure}
 \centering
\begin{tabular}{cc}
\hspace*{-1cm}\includegraphics[width=18cm]{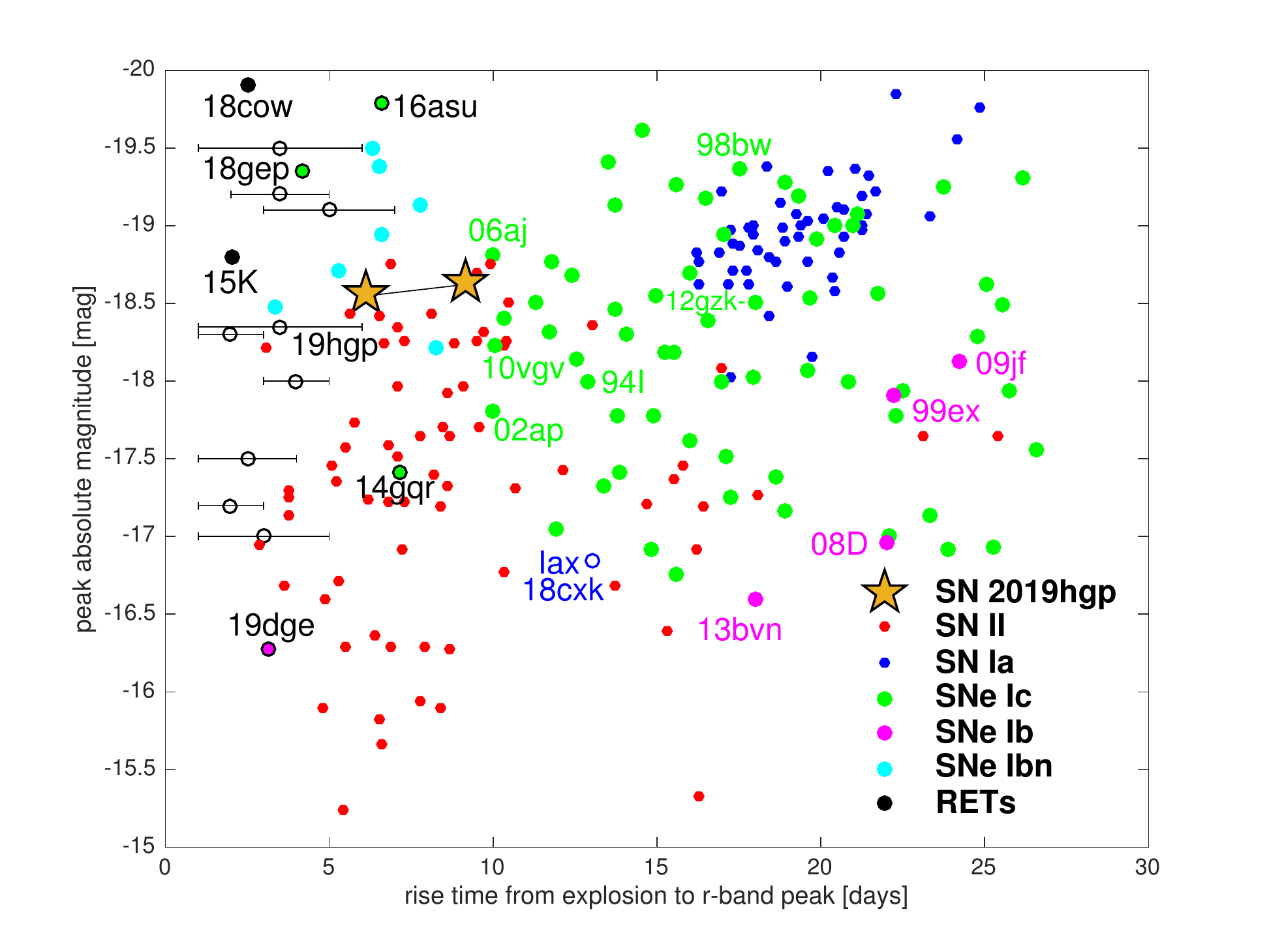}
\vspace*{-2.5cm}
\end{tabular}
\caption{
SN 2019hgp is a rapidly rising, fairly luminous transient. Among H-poor SNe (i.e., excluding Type II SNe, red dots), its location on the rise-luminosity diagram (see Methods $\S~3$ for the rise-time derivation) is similar to those of the Type Ibn events (cyan) and differs from those of all other classes. The object shares a similar phase-space location with well-observed rapidly-evolving transients (RETs, black circles; RETs with spectral similarities to SNe Ic and Ib, green and magenta filled circles, respectively; a sample of RETs that lack spectroscopic classification, open black symbols with duration uncertainty noted as error bars). 
See Methods $\S~10$ for additional details and data sources.
  \label{fig:risetime}}
 \end{figure}

The Zwicky Transient Facility (ZTF)\cite{Bellm2019} first detected SN 2019hgp (ZTF19aayejww) located at J2000 right ascension $\alpha = 15^{h}36^{m}12.86^{s}$ and declination $\delta = 39^{\circ}44'00.5''$ in $r$-band images obtained starting 2019 June 8.2422 UTC, about $1.1$\,d after the estimated explosion time (see Methods $\S~2$). We promptly obtained a spectrum of this object (Fig.~\ref{fig:spec}), which is unique, dominated by highly ionized emission lines of carbon and oxygen, and lacking prominent lines of both hydrogen and helium. Its redshift is consistent with that of the nearby host galaxy ($z=0.0641\pm0.0002$). A rapid follow-up campaign was triggered\cite{Gal-Yam2011} and we collected densely-sampled optical and ultra-violet (UV) photometry and spectroscopy (see Methods $\S~3$ and $\S~6$). The object rapidly rose to maximum brightness in $r$: $<9.5$\,d compared to typically 15\,d for most hydrogen-deficient supernovae (SNe)\cite{Perley2020} (Fig.~\ref{fig:risetime}), placing it among RETs.

\begin{figure}
 \centering
\vspace*{-6.5cm}
\begin{tabular}{cc}
\hspace*{-1.5cm}
\includegraphics[width=17cm]{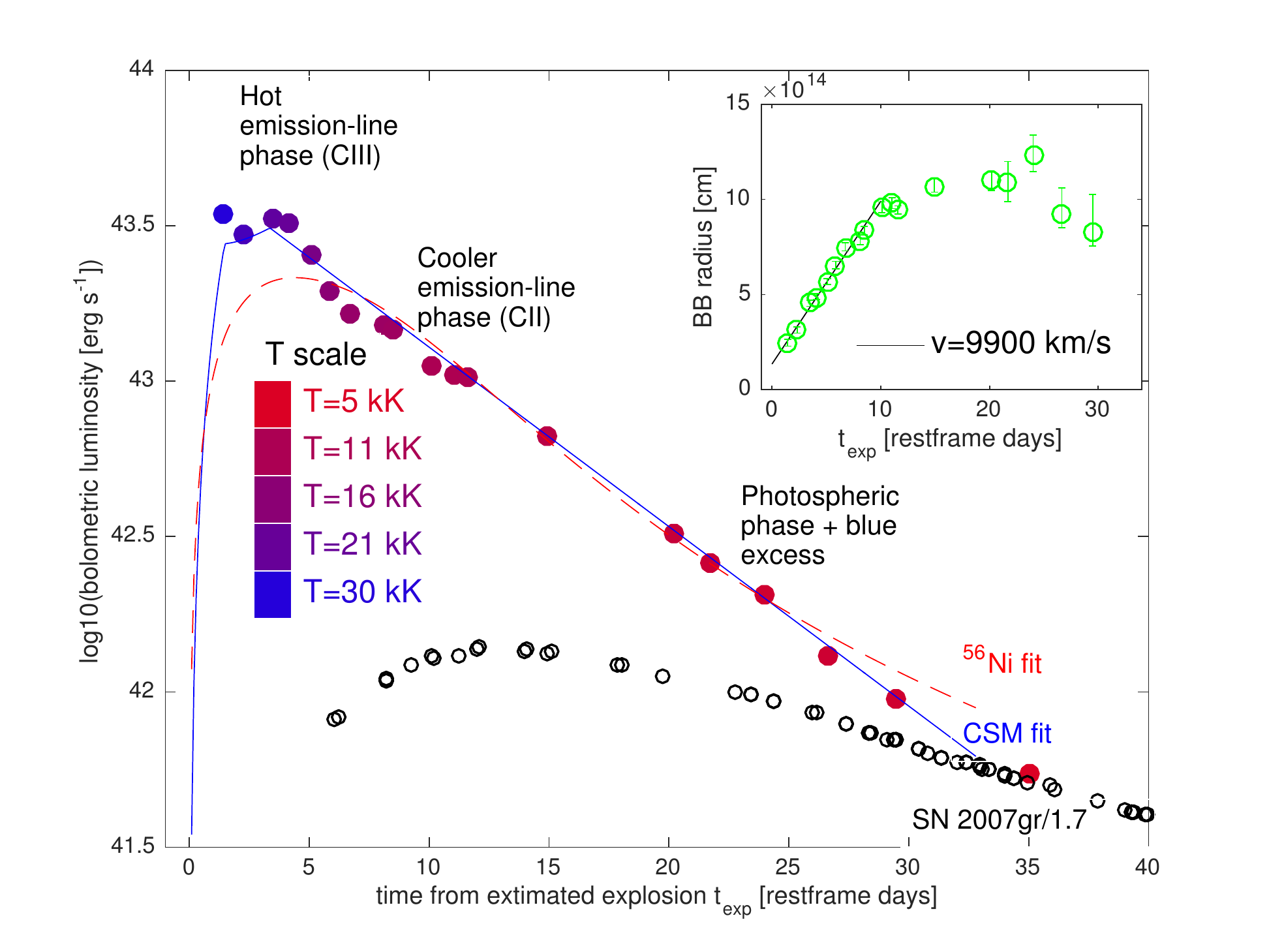}
\vspace*{-17cm}
\end{tabular}
\caption{
The bolometric evolution of SN 2019hgp shows rapid cooling from an initial hot phase. Two weeks post-explosion the spectral energy distribution (SED; Extended Data Fig.~4) is well described by black-body (BB) curves, and the inferred black-body radii (inset) indicate an expansion velocity of $\sim9900$\,km s$^{-1}$. A clear blue excess above the best-fit black-body SED appears around day 15 (Extended Data Fig.~4); black-body parameters (radius and temperature) are less reliable after that date. The light curve is well fit by models of CSM interaction (solid blue); radioactive models (dashed red) cannot fit the peak data even if the entire ejecta are composed of $^{56}$Ni, which is ruled out by the spectra (Extended Data Figures~1-3). A scaled light curve (black) of the well-observed rapidly-declining Type Ic SN 2007gr\cite{Sharon+Kushnir2020} is shown for comparison. Standard 1$\sigma$ error bars marked. 
\label{fig:LC_Bol}}
 \end{figure}

A bolometric light curve derived from our photometry 
(Methods $\S~5$) is plotted 
in Fig.~\ref{fig:LC_Bol}. It demonstrates the vivid contrast between the rapid rise and decline of this event and the much slower evolution of a typical hydrogen-poor SN. Comparing the light curve to models\cite{Kool2020} using \textit{Tigerfit}
(Methods $\S~11$), we find (Fig.~\ref{fig:LC_Bol}) that our early photometric data cannot be explained by models based on energy release from freshly synthesized radioactive $^{56}$Ni,\cite{Arnett1982} as is commonly assumed for H-deficient (Type I) SNe\cite{Drout2011,Taddia2018,Prentice2019}. Instead, simple models based on interaction\cite{Chatzopoulos2012} between the expanding ejecta from the explosion and a distribution of circumstellar material (CSM) fit the data well, and indicate an explosion emitting a total radiated energy of $E_{\rm rad}=0.11\times10^{51}$\,erg, and a compact progenitor with a pre-explosion radius of $R_*=4.1\times10^{11}$\,cm. The properties of the ejecta are a total 
mass of $M_{\rm ej}= 1.2$\,M$_{\odot}$ with an opacity of $\kappa = 0.04$\,cm$^2$\,g$^{-1}$, as expected for C/O mixtures\cite{Rabinak2011}. The total CSM mass required is $M_{\rm CSM}=0.2$\,M$_{\odot}$; the mass-loss rate is $\dot{M}=0.004$\,M$_{\odot}$\,yr$^{-1}$ expanding at a velocity of 
$v_{wind}=1900$\,km\,s$^{-1}$.   

\begin{figure}
\centering
\begin{tabular}{cc}
\hspace*{-3cm}\includegraphics[width=21cm]{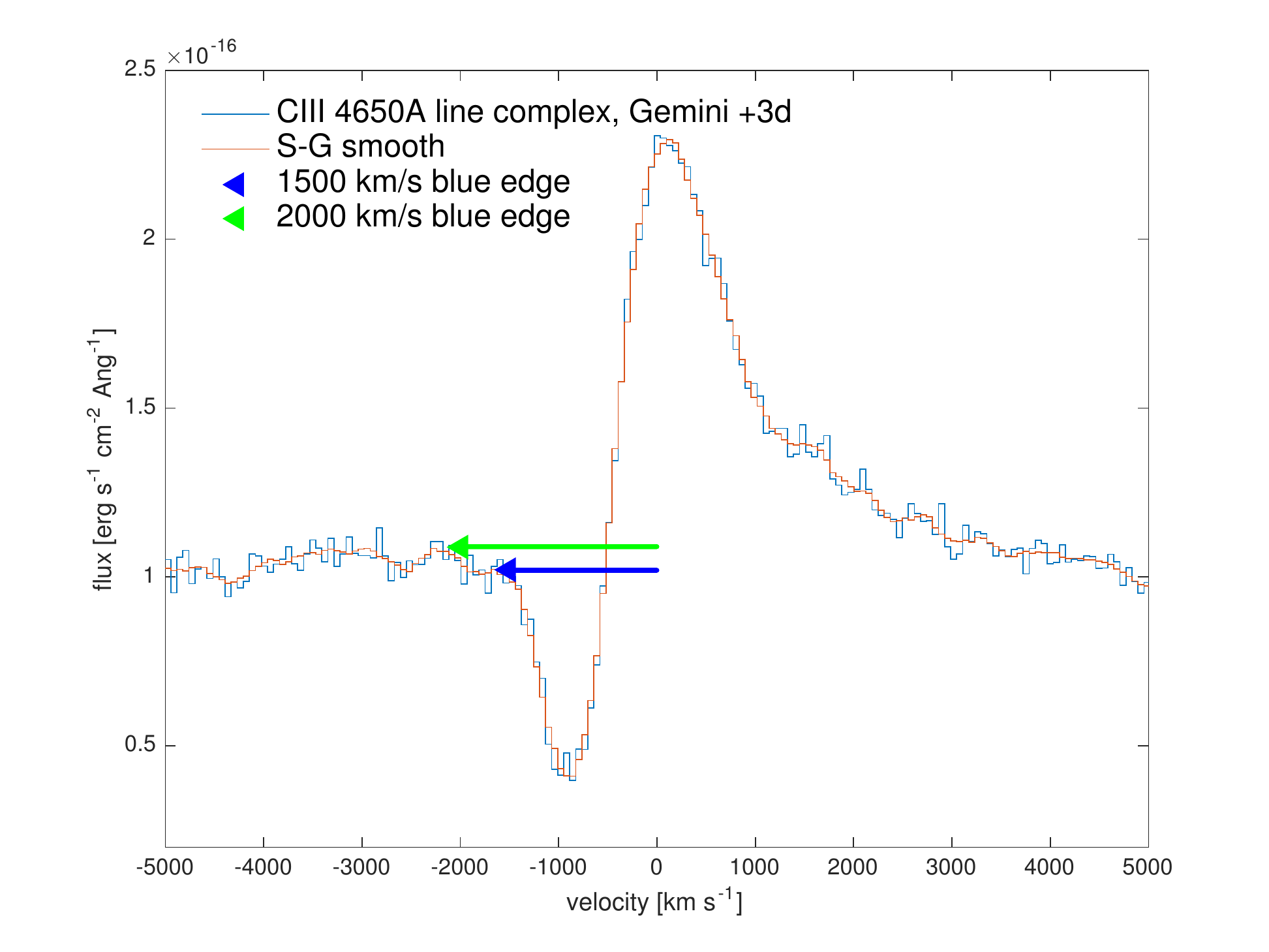}
\vspace*{-3cm}
\end{tabular}
\caption{
SN 2019hgp is embedded in a rapidly-expanding nebula. 
Absorption from twice-ionized carbon in our early-time spectra indicates a line of sight expansion velocity of the nebula surrounding SN 2019hgp with a blue edge extending at least out to $1500$\,km\,s$^{-1}$ (blue), and potentially to $2000$\,km\,s$^{-1}$ (green arrow), considering the uncertainty in the location of the absorbed continuum. High-frequency noise has been smoothed using a third-order Savitzky-Golay (S-G; red) filter. 
  \label{fig:blue_edge}}
 \end{figure}

We obtained an extensive series of spectra of SN 2019hgp (Extended Data Figures~1-3). Our initial data revealed a hitherto unobserved rich set of emission lines that persist for about 20\,d. Line identification shows that these arise from a nebula composed of carbon, oxygen, and neon, with no obvious trace of hydrogen or helium (Fig.~\ref{fig:spec}). We could find no similar spectra among thousands of previously reported observations of explosive transients. Some of the strongest spectral lines present a clear P~Cygni profile, a combination of absorption and emission from an expanding nebula, commonly seen in spectra of massive stars embedded in thick winds. From our best data we measure a wind expansion velocity $>1500$\,km\,s$^{-1}$ (Fig.~\ref{fig:blue_edge}), typical of W-R stars. Observationally, W-R stars are broadly categorized into WN stars (showing strong  spectroscopic features of He, N, and sometimes H) and WC stars (exhibiting features of C and O, but not of H or He). The spectra of SN 2019hgp therefore indicate a CSM nebula similar to those of W-R stars of the WC family. An expansion velocity of $\sim2000$\,km\,s$^{-1}$, as indicated by our light-curve modelling, is consistent with the spectra (Fig.~\ref{fig:blue_edge}).

The final fate of W-R stars is an open problem in astrophysics. Basic considerations suggest that all stars above a cutoff initial mass of 8--10\,M$_{\odot}$, including W-R stars, should at the end of their lives fuse their core material to inert iron and undergo core collapse\cite{Janka2012}. For many years, W-R stars were considered natural candidate progenitor stars for SNe of Types Ib and Ic --- stellar explosions that do not exhibit signatures of hydrogen (Type Ib) or even helium (Type Ic)\cite{Filippenko1997}. However, several lines of evidence suggest that the observed population of SNe~Ib/c cannot arise solely from massive W-R stars\cite{Smith2011a,Taddia2018,Dessart2011}. 

Our observations suggest that SN 2019hgp did arise from an explosion of a massive star that had very similar properties to those of a WC type W-R star. The rapid rise and decline of the light curve imply that the total ejected mass was small ($\sim1$\,M$_{\odot}$ if we adopt the simple CSM model; Methods $\S~11$). If so, a WC progenitor star within the observed mass range of this class (9--16\,M$_{\odot}$)\cite{Crowther2007} suggests that the remnant of the explosion must have been a black hole, as the ejecta are too light to carry the excess mass above that of a neutron star. However, this tentative conclusion is subject to at least the following two caveats. First, a period of enhanced mass loss as indicated from our modelling, with mass-loss rate which is $>100$ times above the typical values for WR stars\cite{Crowther2007}, occurring prior to explosion, may have significantly reduced the pre-explosion total mass of the progenitor star. Second, the ejected mass is estimated using rather simple spherical models (see Methods $\S~11$), and in any case cannot account for ``dark'' mass that cools rapidly after explosion and is not energized by radioactivity or CSM interaction. A combination of such caveats may significantly reduce the apparent gap between the derived ejecta mass and the estimated pre-explosion progenitor mass.

SN 2019hgp is included in the ZTF Bright Transients Survey (BTS)\cite{Fremling2020,Perley2020} and its first spectrum was sufficient to identify its unique nature (Extended Data Fig.~1); we can therefore estimate from having but a single event in this survey that similar events comprise a small fraction of the total core-collapse SN rate, of order $10^{-3}$. 

Of particular interest is the detection of Ne III lines. Such lines have not been observed before in the context of material stripped off of an evolved star (rather than as trace elements within a nearly solar-composition wind). 
The neon observed here was likely the nucleosynthetic product of the same processes that formed the C/O layer, and is therefore likely to be dominated by $^{20}$Ne; further study of these data may illuminate the formation process of cosmic Neon.   

W-R stars of the WN type have previously been proposed as progenitors\cite{Foley2007,Pastorello2008,Karamehmetoglu2017} of a subset of transients (Type Ibn supernovae; \cite{agy17,Pastorello2008,Hosseinzadeh2017}) that, as a class, show the most rapidly evolving light curves among all SNe (Fig.~\ref{fig:risetime})\cite{Perley2020,Hosseinzadeh2017}, and whose spectra indicate the progenitors must have been rich in He (and sometimes also show traces of H)\cite{Pastorello2008,agy17,Pastorello2016,Hosseinzadeh2017}. Combined with our new observations, this suggests an emerging picture where W-R stars can explode as SNe appearing as rapidly evolving transients (rather than typical SNe with longer rise and decline times); WN stars may end their lives as SNe Ibn\cite{Foley2007}, and WC stars may be the progenitors of events like SN 2019hgp\cite{Perley2021a,Pastorello2021} that require a new spectroscopic class --- Type Icn is the obvious choice\cite{Gal-Yam2021}. 

%%%%%%%%%%%%%%%%%%%%%%%%%%%%%%%%%%%%%%%%%%%%%%%%%%%%%%%%%%%%%%%%%%%%%%%%%%%%
%%%%								BIBLIOGRAPHY 1
%%%%%%%%%%%%%%%%%%%%%%%%%%%%%%%%%%%%%%%%%%%%%%%%%%%%%%%%%%%%%%%%%%%%%%%%%%%%
%%%
\clearpage

%ADS custom format: \\bibitem{%2h%Y}\n\bibinfo{author}{%1G}\n\newblock \bibinfo{title}{%T}.\n\newblock \emph{\bibinfo{journal}{%J}} \textbf{\bibinfo{volume}{%V}},\n\bibinfo{pages}{%p} (\bibinfo{year}{%Y}).

%%%%%%%%%

%%%%%%%%%%%%%%%%%%%%%%%%%%%%%%%%%%%%%%%%%%%%%%%%%%%%%%%%%%%%%%%%%%%%%%%%%
%								FIGURES
%%%%%%%%%%%%%%%%%%%%%%%%%%%%%%%%%%%%%%%%%%%%%%%%%%%%%%%%%%%%%%%%%%%%%%%%%

 \clearpage

 \clearpage
 
\clearpage

\clearpage

 %%%%%%%%%%%%%%%%%%%%%%%%%%%%%%%%%%%%%%%%%%%%%%%%%%%%%%%%%%%%%%%%%%%%%%%%%
%								METHODS
%%%%%%%%%%%%%%%%%%%%%%%%%%%%%%%%%%%%%%%%%%%%%%%%%%%%%%%%%%%%%%%%%%%%%%%%%

\clearpage

\begin{methods}

\subsection{1. Summary of Observations:}

Our photometric observations are provided in Online Supplementary Table~\ref{phot_tab}, shown in Extended Data Fig.~5, and discussed in the Methods Section $\S~3$ below. Spectroscopic data are presented in Extended Data Figures~1-3, and details about the observational setups are provided in Supplementary Table~\ref{spec_tab} and Methods Section $\S~6$.

\subsection{2. Detection of SN 2019hgp and its estimated explosion time:}

SN 2019hgp was first detected by ZTF\cite{Bellm2019,Graham2019} 
located at J2000 right ascension $\alpha = 15^{h}36^{m}12.86^{s}$ and declination $\delta = 39^{\circ}44'00.5''$, with an estimated positional uncertainty of $0.44''$ compared to {\it Gaia}\cite{Yaron2019}, 
in an $r$-band image obtained with the ZTF camera\cite{Dekany2020} on JD 2,458,642.7422 (2019 June 8.2422 UTC), about $1$\,d after the last nondetection by the same instrument. The ZTF image-processing pipeline\cite{Masci2019} generated an alert based on image subtraction\cite{Zackay2016} with respect to a reference image. The alert was picked up by our custom ``infant supernovae'' filter\cite{Gal-Yam2019} running on the ZTF Growth Marshal system\cite{Kasliwal2019}. It was identified by a duty astronomer (R.J.B.) and follow-up observations were promptly triggered using our standard methodology\cite{Gal-Yam2011}. The object was reported to the IAU Transient Name Server (TNS; \url{https://wis-tns.weizmann.ac.il/object/2019hgp}) on June 10, 2019\cite{Bruch2019} and was allocated the name AT 2019hgp.  Forced photometry analysis performed at the SN location using custom methodology\cite{Strotjohann2021} recovered prediscovery signal in stacked $r$-band images obtained during the night prior to discovery (Extended Data Fig.~6).   

To estimate the explosion time, we fit low-degree polynomials to our well-observed $r$-band light curve and adopt the mean and standard deviation of these fits (2019 June $7.1\pm0.2$ UTC) as an estimate for the explosion time and its uncertainty (Extended Data Fig.~6). All times reported in this paper are with respect to this explosion time t$_{\rm exp}$.  

\subsection{3. Photometry:}
ZTF $gri$ photometry obtained with the ZTF survey camera was processed with the ZTF image reduction pipeline\cite{Masci2019} employing the ZOGY image-subtraction method\cite{Zackay2016}. We obtained additional $ugri$ photometry with the robotic 60-inch telescope at Palomar (P60\cite{cfm+06}), using the Spectral Energy Distribution Machine (SEDM\cite{Blagorodnova2018}), extracting PSF photometry from image subtraction against SDSS templates using {\tt FPipe}\cite{fst+16}. Additional $ugriz$ photometry was obtained using the IO:O camera mounted on the 2m Liverpool Telescope (LT) and reduced using the telescope standard software followed by our custom subtraction pipeline based on {\tt FPipe}.  

Photometry was also obtained using 
%Gemini/GMOS and 
GTC/OSIRIS as part of our spectroscopic campaign.  
%We reduced the Gemini acquisition images using the {\tt GEMINI} IRAF package that uses standard routines in IRAF \cite{Tody1986a} to de-bias and flat-field the images. We then solved the astrometry using stars from GAIA DR2 using the software package {\tt Gaia} version 4.4.6 (\url{http://starlink.eao.hawaii.edu/starlink/2015ADownload}).
We obtained GTC photometry in $g'$, $r'$, $i'$ and $z'$ during the same night. The data were analysed using the OSIRIS Offline Pipeline Software ({\tt OOPS}) version 1.4.5 (\url{http://gtc-osiris.blogspot.com/2012/10/the-osiris-offline-pipeline-software.html}) that employs standard routines in IRAF to de-bias and flat-field the images. 
We then solved the astrometry using stars from GAIA DR2 using the software package {\tt Gaia} version 4.4.6 (\url{http://starlink.eao.hawaii.edu/starlink/2015ADownload}).
%We solved the astrometry as we did for the Gemini data. 
We performed aperture photometry using a custom tool\cite{Schulze2018} available from \url{https://github.com/steveschulze/Photometry}. Once an instrumental magnitude was established, it was photometrically calibrated against the brightness of several standard stars measured in a similar manner and tied to the SDSS DR8\cite{Aihara2011}. 

SN 2019hgp was also observed with the Ultra-Violet/Optical Telescope (UVOT\cite{Roming2005}) on board the Neil Gehrels \textit{Swift} Observatory\cite{Gehrels2004}. Observations began 8.1 hours after the ZTF discovery.
The UVOT data were retrieved from the NASA \textit{Swift} Data Archive (available at \url{https://heasarc.gsfc.nasa.gov/cgi-bin/W3Browse/swift.pl}) and reduced using standard software distributed with {\tt HEAsoft} version 6.19 (available at \url{https://heasarc.nasa.gov/lheasoft/}), using the recently revised calibration.
Photometry was measured using {\tt uvotmaghist} with a $3''$ radius circular aperture. To remove the host contribution, we obtained a final epoch in $B$ and $V$ on 2 and 3 August 2019 and used archival data in $w2$, $m2$, $w1$, and $U$ that were obtained between 2007 and 2008. We built a host template using {\tt uvotimsum} and {\tt uvotsource} with the same aperture used for the transient. We then numerically subtracted the host flux from the transient light curve.

Extended Data Fig.~5 shows the observed light curves, and all photometry is listed in Supplementary~Table~\ref{phot_tab}. All photometry has been corrected for Milky Way foreground extinction according to $E(B-V) = 0.019$\,mag\cite{sf11}, and assuming a negligible host extinction (Methods $\S~5$). We assume a standard $\Lambda$CDM 
cosmology\cite{ksd+11}
with $\Omega_{\rm M} = 0.27$, $\Omega_{\Lambda} = 0.73$ and $H_0 = 70$~km~s$^{-1}$~Mpc$^{-1}$.
%\cite{ksd+11}. 

\paragraph{Rise time:} Estimation of the object $r$-band rise time is complicated as the light curve shows low-level undulations around peak. To estimate the $r$-band peak date we use what we consider to be our most reliable data set (forced-photometry observations from the ZTF P48 wide-field camera), binned to 1-day bins. These observations show two apparent peaks at $6.15$\,d and $9.15$\,d restframe days from our estimated explosion date, with the second peak being slightly more luminous ($-18.58$\,mag and $-18.64$\,mag, respectively). Smooth rise and decline precede and follow these two peaks so we consider the date of the peak to be securely within this range. We plot both values in Fig.~\ref{fig:risetime}.

\subsection{4. Pre-explosion limits:}
In addition to the limits from the supernova discovery observing season listed in Supplementary~Table~1, the field was observed by PTF, iPTF and ZTF a number of times prior to the discovery of the supernova. Pre-explosion limits exist for the following date ranges: May 13-19 and July 5, 2009; March 18 - June 13, 2010; March 1-2, 2011; February 1 - June 20, 2013; March 19 - May 28 and December 20, 2014; and June 5-26, 2015. All upper limits are in $r$-band except for March 1-2, 2011 and June 5-26, 2015 which are $g$-band. Typical nightly upper limits are between 20.5 - 21.5~mag, i.e., constraining pre-explosion eruptions with peak absolute magnitudes of $-17.5$ to $-16.5$\,mag. 
Within the 2.5 years before the SN explosion ZTF observed its position 915 times in 194 different nights. When combining observations in 7-day-long bins\cite{Strotjohann2021} we can rule out eruptions brighter than an absolute magnitude of -15.5 in the g or r band during $20\%$ of the duration of ZTF (corresponding to $56\%$ of the periods with observations).
As these limits are brighter than some of the precursors detected prior to SNe of Types IIn and Ibn so far\cite{Pastorello2007,Ofek2013,Ofek2014a,Strotjohann2021}, we cannot put strong constraints on the eruptive history of the progenitor of SN 2019hgp.

\subsection{5. Spectral energy distribution evolution and extinction:}

Using our well-sampled photometry of SN 2019hgp extending from the UV to the near IR (Supplementary Table~\ref{phot_tab}; Extended Data Fig.~5), we can trace the spectral energy distribution of the event (Extended Data Fig.~4) and its evolution with time, and construct the bolometric light curve (Fig.~\ref{fig:LC_Bol}). To calculate the bolometric light curve from our UV-IR photometry, we fit a black body (BB) curve to each epoch, and integrate the flux enclosed within the wavelength range covered by our photometry (typically extending from the {\it Swift} UVW2 to the $z$-band). During epochs where the data is well fit by a black body (BB), we adopt the total integrated BB luminosity as the bolometric value. In later epochs we detect a UV excess above the BB curves (Extended Data Fig.~4); and we therefore estimate the bolometric luminosity using the integrated observed flux with UV and IR corrections calculated by integrating under the BB curves outside of the range covered by our data. We note that any UV excess above the BB curve blueward of the {\it Swift} bluest band (UVW2) cannot be accounted for, and in these later epochs our adopted bolometric values are therefore lower limits.

The bolometric evolution of SN 2019hgp shows rapid cooling from an initial very hot phase (T$\approx30$~kK; Fig.~\ref{fig:LC_Bol}), rarely observed before. 
During the initial two weeks, the spectral energy distribution (SED; Extended Data Fig.~4) is well described by black body curves with temperatures cooling to $\sim10$\,kK on day 15; the inferred black body radii (Fig.~\ref{fig:LC_Bol}; inset) indicate a photospheric expansion velocity of
$\sim9900$ km s$^{-1}$. The appearance of a UV excess beyond this time, as well as the spectroscopic evolution (Methods~\S~9; Extended Data Fig.~7), all suggest a dominant contribution from interaction during the later phases, with the relevant CSM located at radii $>10^{15}$\,cm.  

\paragraph{Extinction:}
SN 2019hgp exploded at the outskirts of its host (Extended Data Fig.~8; see below Methods $\S~8$), does not show evidence for narrow Na D absorption at any phase, and is initially very blue, indicating that the host extinction of this object is unlikely to be significant. We therefore assume throughout the paper no extinction at the host.

We can use our early data to investigate the range of allowed extinction values. Fitting our first epoch optical-UV SED, we find that the data are well fit assuming negligible extinction, but allow higher extinction values (indicating of course a higher BB temperature; Extended Data Fig.~9). Regardless of the extinction law parameters (Galactic, LMC or SMC curves, and the value of R$_{\rm V}$), the maximal extinction values allowed are E$_{\rm B-V}\sim0.15$ requiring an initial temperature as high as $\sim100$\,kK (Extended Data Fig.~9).

\subsection{6. Spectroscopy:}
We obtained a total of 33 spectra of SN 2019hgp, taken with the instruments listed in Supplementary~Table~2. 
The sequence of spectra is shown in Extended Data Fig.~1-3. All spectra will be made publicly available through the Weizmann Interactive Supernova Data Repository (WISeREP\cite{yg12,WISeREP}).

\paragraph{P60/SEDM:} The Spectral Energy Distribution Machine (SEDM \cite{Ben-Ami2012a,Blagorodnova2018}) is an integral field unit spectrograph with a low resolution of $R\sim100$ mounted on the 60" robotic telescope (P60 \cite{cfm+06}) at Palomar observatory. It is primarily used to rapidly vet SN candidates discovered by the ZTF survey and the first spectrum of SN\,2019hgp was obtained by the SEDm only 4.3 hours after the SN was detected. SEDM data are reduced automatically \cite{Rigault2019}.

\paragraph{GMOS/Gemini:} After the initial SEDM spectrum, a higher resolution spectrum was obtained with the Gemini Multi-Object Spectrograph (GMOS; \cite{Hook2004}) mounted on the Gemini North 8m telescope at the Gemini Observatory on Mauna Kea, Hawaii. Two 900\,s exposures were obtained with the B600 grating and with central wavelengths of 520\,nm and 525\,nm, respectively, to cover the chip gap. The same setup was used for the second Gemini spectrum on 2019 June 10th. The GMOS data were reduced using the Gemini IRAF package version 1.1.14.

\paragraph{LT/SPRAT:} The Spectrograph for the Rapid Acquisition of Transients (SPRAT; \cite{Piascik2014}) is a high-throughput, low-resolution spectrograph mounted on the Liverpool Telescope (LT; \cite{Steele2004}), a 2 meter robotic telescope at the Observatorio del Roque de Los Muchachos in Spain. LT spectra of SN\,2019hgp were reduced using the standard pipeline provided by the observatory.

\paragraph{NOT/ALFOSC:} We observed the object with the Alhambra Faint Object Spectrograph and Camera (ALFOSC) mounted on the 2.56\,m Nordic Optical Telescope (NOT) based at the Roque de los Muchachos Observatory. The spectra were reduced in a standard way, which includes wavelength calibration through an arc lamp, and flux calibration utilizing a spectrophotometric standard star.

\paragraph{HET/LRS2:} We also obtained optical spectra of SN\,2019hgp with the Low Resolution Spectrograph 2 (LRS2\cite{Chonis2016}) on the 10-meter {\sl Hobby-Eberly Telescope}\cite{Ramsey1998}.
LRS2 has blue (LRS2-B) and red (LRS2-R) arms; each arm is a dual-arm spectrograph. The UV and orange arms on LRS2-B cover the spectral ranges of 3700$-$4700 \AA\ with a resolving power of $R\sim$1900, and 4600$-$7000 \AA\ with $R\sim1100$, respectively. The two arms of LRS2-R cover 6500$-$8420 \AA\ and 8180$-$10500 \AA, both with a spectral resolving power of $R\sim$1800. Each arm is fed by separate $12\arcsec \times 6\arcsec$ integral field units (IFU)\cite{Chonis2016}. The three first HET spectra of SN\,2019hgp were obtained with the blue arm only, while both arms were used sequentially for the two later spectra. The red arm data were not useful during the last epoch.

The LRS2 IFU data were reduced with self-developed IRAF and Python scripts. Fiber-to-fiber transmission variations were corrected with twilight flat-field frames obtained during the same night. Spectra obtained with the LRS2-B and LRS2-R were wavelength calibrated based on the spectra of HgCd and FeAr lamps, respectively. For each epoch of observation, a mean sky spectrum was constructed by median combining the flux of all fibers after a 3$\sigma$-clipping procedure. Flux calibration was carried out each night by observing spectrophotometric standard stars at similar airmasses. Finally, we corrected for the telluric lines using a mean spectrum constructed from observations of telluric standard stars.

\paragraph{WHT/ACAM:} One spectrum was obtained with the single slit Auxiliary-port CAMera spectrograph (ACAM\cite{Benn2008}) mounted on the 4.2m William Herschel Telescope (WHT) at the Observatorio del Roque de los Muchachos in La Palma, Spain. The ToO was obtained as part of the Optical Infrared Coordination Network for Astronomy (OPTICON) program. The spectrograph has an approximate resolution of $R\sim400$ and spectral data were reduced using standard IRAF routines.

\paragraph{LDT/Deveny/LMI:}
Spectroscopy was obtained with the DeVeny Spectrograph on the 4.3\,m Lowell Discovery Telescope in Happy Jack, Arizona (LDT, formerly the Discovery Channel Telescope or DCT\cite{Levine2012,Levine2016}) on 2019 June 22. The LDT spectrum (PI: Gezari) was obtained with a 1.5" wide slit and taken in two 450 second exposures with the 300\,g/mm grating. We reduced the spectrum with standard IRAF routines, stacking the exposures into a single 2D science frame, and corrected for bias and flat-field before extracting the 1D spectrum. The spectrum was wavelength calibrated by comparing with spectra of HgNeCdAr arc lamps, and flux calibration was performed using the standard star Feige 67.

\paragraph{P200/DBSP:}
The Double Beam SPectrograph (DBSP\cite{Oke1982}) is mounted on the 5m Hale telescope at Palomar Observatory (P200). The two spectra were obtained with a 600/4000 grism on the blue side and a 316/7150 grating on the red side yielding a spectral resolution of $R\sim 1000$. The data were reduced with the pyraf-dbsp pipeline\cite{Bellm2016}.

\paragraph{Keck/LRIS:} Two spectra of the fading SN were obtained with the Low-Resolution Imaging Spectrometer (LRIS\cite{Oke1995}) mounted on the Keck-I 10m telescope at the W. M. Keck Observatory in Hawaii. The data were reduced with the LRIS automated reduction pipeline Lpipe\cite{Perley2019a}.

\paragraph{GTC/OSIRIS:} We used the 10.4\,m Gran Telescopio Canarias (GTC), situated on the island of La Palma, Spain, to obtain late-time spectroscopy of SN\,2019hgp. Director Discretionary Access to the facility was most kindly granted and proved critical as at that time all facilities on top of Mauna Kea were shut down.
The spectra were obtained with the OSIRIS instrument (Optical System for Imaging and low-Intermediate-Resolution Integrated Spectroscopy) using the grisms R1000B and R1000R, with an exposure time of $3 \times 1400$s in each grism. The observations with the two arms were performed in two consecutive nights (29 and 30 July, respectively) and the spectra were co-added to produce a single spectrum covering the wavelength range 3600 -- 10200 \AA. All spectra were reduced and calibrated using custom-made pipelines, based on IRAF. 

\paragraph{Redshift:}

During our first Gemini observations ($1.4$\,d after explosion; Supplementary Table~\ref{spec_tab}) we extracted a spectrum of both the transient and the nearby potential host galaxy, for which there was no catalogued redshift information. We measure a host redshift of $z_{\rm host}=0.0641\pm0.00001$, where the error represents only the statistical error from the scatter of values obtained from fitting individual strong lines (H$\beta$, OIII$\lambda\lambda$4959,5007\AA, HeI $\lambda$5876\AA), weighted by the line error measurements. Measuring the transient redshift from the same data using the strongest isolated lines of CIII ($\lambda5696$\AA) and OIII ($\lambda5007$\AA) we find a value of 
$z_{\rm transient}=0.0638\pm0.00001$.
Comparing the transient redshift values measured from Gemini data  obtained on two different epochs (1 and 3\,days after explosion), we estimate these values have an additional systematic uncertainty of $\Delta z=0.0002$. The measured velocity offset between the transient and its host ($\Delta z=0.0003$; v$=90$\,km\,s$^{-1}$) is well within the velocity distributions of stars within galaxies. Since the transient is also superposed on a diffuse component of the apparent host (Extended Data Fig.~8), we consider the association of the transient with the host to be secure. Since the transient emission might be shifted by the intrinsic bulk velocity of the expanding material, we adopt the host redshift when we calculate the distance to this event; the slight offsets above have in any case negligible impact on our calculated results.    

\paragraph{Early emission-line phase:}

The early spectra of SN 2019hgp (Days $1-6$, Extended Data Fig.~1) show a hot, blue continuum consistent with the hot black-body fits (Extended Data Fig.~4) on which numerous emission lines are superposed. Analysis of our high-resolution spectra (Fig.~\ref{fig:spec}) show that the emission is dominated by highly ionized carbon, oxygen and neon. Helium (or hydrogen) lines are not obvious in any spectrum, making this object remarkably different from any previously-observed transients. 

The dominance of carbon and oxygen, along with the low expansion velocity determined from the P Cygni absorption features (Fig.~\ref{fig:blue_edge}), which is significantly below the photospheric expansion velocity estimated from our BB fits (Fig.~\ref{fig:LC_Bol}; inset) suggests the emission lines come from a unique distribution of CSM surrounding the exploding star. The apparent composition, lacking strong lines of hydrogen and helium, is similar to that expected from Wolf-Rayet stars of types WC and WO. 

Focussing on the presence of He II in particular, we note that the peak of the emission bump seen near the location of He II $\lambda5411$\AA\ is offset by about $10$\AA\ with respect to the expected wavelength, making the association of this feature with He II uncertain. The strongest line of He II in the visible range, $\lambda$4686\AA, is blended with the red wing of the strong CIII line at $\lambda$4650\AA. To further test whether He II contributed to this area of the spectrum, we modelled the observed spectrum in this region with a P Cygni profile of CIII, composed of a Lorenzian emission profile with a blueshifted Gaussian absorption feature (Extended Data Fig.~10). We then tested whether introducing an additional Lorentzian emission component at the wavelength of He II $\lambda$4686\AA\ is favored in a $\chi^2$ sense. Our analysis indicates that this is indeed the case (Extended Data Fig.~10), but that both models provide a reasonable description of the data. We conclude therefore that while our spectra do not show obvious evidence for He II emission, the presence of this ion is permitted by our observations. This is consistent with the spectroscopic analysis of WC stars\cite{Sander2012} where models that include $55\%$ He by mass fit the spectra well, with only marginal emission from He II $\lambda$5411\AA\ and with the He II $\lambda$4686\AA\ blended into the strong CIII complex, as we see.
As for the presence of hydrogen, spectroscopic series of hydrogen-rich supernovae of Type II\cite{Gal-Yam2014,Khazov2016,Bruch2020} show ubiquitous strong emission lines of hydrogen. For example, observations of SN 2013fs\cite{Yaron2017} covering a very broad range of temperatures and obtained during similar phases after explosion, always show strong H$\alpha$ emission. We therefore consider it unlikely that there is hydrogen in this event.    

\paragraph{Late emission-line phase:}

About six days after explosion, the strong emission lines of CIII and OIII have largely disappeared (Extended Data Fig.~2; top) and a set of emission lines of lower ionization species appear, initially of CII and later, around day 10, of OI. One would expect the oxygen population to go through a phase dominated by OII, and indeed a feature reminiscent of the W-shaped OII complex seen in SLSNe-I\cite{agy19} is seen in the P60 +6.9\,d spectrum. However, higher resolution spectra obtained before and after that spectrum resolve those features into residual absorption from OIII and CIII. It would therefore seem that SN 2019hgp did not go through an OII-dominated phase. The spectra obtained around $12-15$\,days are quite featureless, although of lower signal to noise. By day 19.3 (Extended Data Fig.~2, bottom), strong, broad photospheric features appear, marking the transition of the object into the photospheric phase.   

\paragraph{Photospheric phase:}

At 19.3\,d post explosion, broad features emerge with P Cygni profiles (Extended Data Fig.~3). The implied velocities are noticeably higher than previously seen, e.g., the strong OI $\lambda7774$\AA\ line shows a two-component absorption structure with the narrow and  broad components showing blue edges extending to $\sim2500$\,km\,s$^{-1}$ (similar to previously seen line velocities; Fig.~\ref{fig:blue_edge}) and $\sim12000$\,km\,s$^{-1}$, respectively. The velocities of the emerging broad components are similar to those deduced from the photospheric expansion (Fig.~\ref{fig:LC_Bol}; inset). Initially, sharp, narrow emission spikes of CI, CII and OI are superposed on the broader features, but those disappear by day 27.4 (Extended Data Fig.~3, bottom) and the spectrum evolves to resemble that of spectroscopically normal Type Ic SNe around peak. Comparison with a spectrum of SN 2017gr around peak\cite{Valenti2008} (Extended Data Fig.~3; bottom) shows that most line features agree, but several differences are also apparent, especially in the area $5000-7500$\AA. Many of the line features of SN 2019hgp are noticeably narrower, and in some case much weaker (e.g., the Ca II H+K feature).             

To test the contribution to the spectrum from He I lines (and thus the spectroscopic classification of the object) we undertake modelling of the 27.4\,d spectrum using the SYNOW\cite{Branch2005} code, our results are shown in Extended Data Fig.~11.
As can be seen there, this analysis does not support the contribution of He I to the spectrum, suggesting a late-time classification of SN Ic for this object, as also indicated by the similarity to SN 2007gr. We stress that we use SYNOW modelling for line identification and verification only, given the many simplifying assumptions underlying this code, such as spherical, homologous expansion and resonant scattering line formation above a sharp photosphere that emits a blackbody spectrum\cite{Branch2005}. In particular, elemental abundances or relative mass fractions cannot be determined using this approach. 

Recent analysis of SNe Ibn\cite{Karamehmetoglu2019} suggests that the emission and absorption P Cygni components of He I transitions can vary with time and depend on the physical properties of the emitting gas. It may require a more sophisticated modelling to determine how much helium is allowed by the spectra we have obtained, however, we note that the reported analysis\cite{Karamehmetoglu2019} shows that for the transition in question ($\lambda6678$\AA), the emission component, which we do not observe, grows stronger with time. We conclude that our data do not present strong evidence for helium during the photospheric phase.    

\paragraph{Nebular spectrum:}

We have attempted to obtain a nebular spectrum of this rapidly fading transient 52.8\,d after explosion using the GTC. The object was very faint at this time (Extended Data Fig.~5) and the object was setting, limiting the duration of our exposures. We have been able to extract the signal from the combined exposures spanning the wavelength range shown in Supplementary Fig.~\ref{fig:neb}; areas outside of this range are very strongly affected by sky lines. The spectrum shows several broad emission features (e.g., a velocity width of $10,000$\,km\,s$^{-1}$ for Na I D) that coincide with commonly observed nebular lines of Ca, Mg, Na and O. Narrow H$\alpha$ from the underlying host is also observed. Weak absorption features are still apparent for Na I D and Mg I] $\lambda4571$\AA\ suggesting perhaps that the emission is not purely nebular. 

\subsection{7. X-ray observations:}

We monitored the field with the \textit{Swift} X-ray telescope (XRT\cite{Burrows2005}) concurrently with the UVOT observations. We built \textit{Swift}/XRT data products using the {\tt Build XRT Products} web service at \url{http://www.swift.ac.uk/user_objects} which employs the methods described in \cite{Evans2007,Evans2009}. The count-rate light curve was built using the binning modes
``Time" and ``Counts" with default parameters.

Swift XRT recorded no X-ray emission during the entire campaign. The 3$\sigma$ limit on the count rate for the entire period is $6.1\times10^{-4}~{\rm ct/s}$. The count-rate limits on the individual epochs are $\sim0.11~{\rm ct/s}$ in 100-s bins. We used {\tt WebPIMMS} \url{https://heasarc.gsfc.nasa.gov/cgi-bin/Tools/w3pimms/w3pimms.pl} to convert the count-rate limit of the stacked data into a flux limit. Assuming synchrotron radiation with a photon index of 2, a Galactic absorption of $N_X({\rm H}) = 1.61 \times 10^{20}\,{\rm cm}^{-2}$ from\cite{HI4PI2016} and no host absorption, the absorption-corrected flux is $<2.2 \times 10^{-14}\,{\rm erg\,cm}^{-2}\,{\rm s}^{-1}$ between 0.3 and 10 keV for the entire period. This corresponds to a luminosity of $<2.2 \times 10^{41}\,{\rm erg\,s}^{-1}$ between 0.3 and 10~keV at z = 0.0641.

We compare our X-ray data to similar observations of other RETs. We use a recently presented sample\cite{Ho2020} and augment it with observations of CSS161010\cite{Coppejans2020}, iPTF14gqr\cite{De2018},
SN 2019dge\cite{Yao2020}, SN 2018gep\cite{Ho2019} and AT2020xnd (ZTF20acigmel)\cite{Perley2021b}. All {\it Swift} XRT data are analyzed as detailed above, while Chandra observations of CSS161010 are converted to the same scale assuming a power law spectrum with index 2. Our results are plotted in Supplementary Fig.~\ref{fig:Xsample}. Only two objects (CSS161010 and AT2018cow) are detected in X-rays. Our observations would have detected an X-ray emission similar to that of AT2018cow from SN 2019hgp (and several other RETs), but the sensitivity of the observations and the range of observing time is such that we cannot exclude that any other RET in our sample, including SN 2019hgp, has a similar X-ray luminosity to that of CSS161010.

In the context of interacting SNe, our upper limits constrain the X-ray luminosity to be $1-3$ orders of magnitude below the bolometric peak (lying initially in the UV and moving through the visible toward the IR with time). Such a ratio of X-ray to optical/UV luminosity was measured for other interacting SNe (e.g., Type IIn SN 2010jl\cite{Ofek2014b,Chandra2015} and Type Ibn SN 2006jc\cite{Immler2008}) where the X-rays were actually detected. Lacking a standard comprehensive model for SN CSM interaction it is difficult to provide additional interpretation of the X-ray data without custom modelling which is beyond the scope of this work, except to say that variants of literature models that fit other interacting events could also be applicable for SN 2019hgp.  

\subsection{8. Host Galaxy:}

SN 2019hgp exploded next to an anonymous star-forming galaxy designated as WISEA J153613.08+394357.2 in the NASA Extragalactic Database (NED). As shown in Extended Data Fig.~8, the SN exploded on top of a diffuse extension of the main body of the galaxy, possibly a spiral arm.

We retrieved science-ready coadded images from the \textit{Galaxy Evolution Explorer} (GALEX) general release 6/7\cite{Martin2005}, the Sloan Digital Sky Survey data release 9 (SDSS DR 9\cite{Ahn2012}), and preprocessed Wide-field Infrared Survey Explorer (WISE) images \cite{Wright2010} from the unWISE archive\cite{Lang2014}. The unWISE images are based on the public WISE data and include images from the ongoing NEOWISE-Reactivation mission R3\cite{Mainzer2014, Meisner2017}. In addition to this, we use the UVOT observations that were obtained either before the explosion of SN 2019hgp or after the SN faded. The brightness in the UVOT filters was measured with UVOT-specific tools in the HEAsoft version 6.26.1. Source counts were extracted from the images using a region of $10''$. The background was estimated using two circular regions with a radius of $20''$ each close to the SN position. The count rates were obtained from the images using the {\it Swift} tool uvotsource. They were converted to magnitudes using the UVOT calibration file from September 2020. All magnitudes were then transformed into the AB system\cite{Breeveld2011}.

We measured the brightness of the host using the Lambda Adaptive Multi-Band Deblending Algorithm in R (LAMBDAR)\cite{Wright2016,Lambdar} and the methods described in\cite{Schulze2020}. The brightness of the host in the UVOT images was measured with the {\it Swift} FTool {\tt uvotsource} using an aperture encircling the entire galaxy. Supplementary Table~\ref{tab:hostphot} details the measurements in the different bands. 

We modelled the host spectral energy distribution with the software package prospector version 0.3 \cite{Leja2017}. Prospector uses the Flexible Stellar Population Synthesis (FSPS) code \cite{Conroy2009} to generate the underlying physical model and python-fsps \cite{Foreman-Mackey2014} to interface with FSPS in python. The FSPS code also accounts for the contribution from the diffuse gas (e.g., HII regions) based on Cloudy models\cite{Byler2017}. We assumed a Chabrier initial mass function\cite{Chabrier2003} and approximated the star formation history (SFH) by a linearly increasing SFH at early times followed by an exponential decline at late times (functional form $t \times \exp\left(-t/\tau\right)$), as well as dust attenuation\cite{Calzetti2000}. Finally, we use the dynamic nested sampling package dynesty \cite{Speagle2020} to sample the posterior probability function.

Supplementary Figure~\ref{fig:gal_sed} shows the observed SED and its best fit. The SED is adequately described by a galaxy template with a mass of
$\log\,M/M_\odot = 9.05^{+0.13}_{-0.24}$ and a star-formation rate of
$0.24^{+0.08}_{-0.04}~M_\odot/{\rm yr}^{-1}$. The mass and the star-formation rate are below average, but still within the distribution of values for host galaxies of Type Ic SNe from the PTF survey\cite{Schulze2020}. SN 2019hgp is located $3.54$'' from the center of its host galaxy. At a redshift of $z=0.0641$ and assuming our adopted cosmology, the offset translates to a projected distance of 4.4 kpc. Although the SN is located in the outskirts of its host, the location is not unusual for Type Ic SNe exploding in galaxies of similar mass \cite{Schulze2020}. 

Type Ic and Type II SNe from the PTF sample exploded in overall similar galaxies\cite{Schulze2020} and have also comparable redshift distributions. This motivates a comparison of the host of SN 2019hgp to those of Type II SNe with similar early CSM signatures (``flash'' features\cite{Gal-Yam2014}). Supplementary Figure~\ref{fig:flash-hosts} presents a kernel density estimate of the host galaxy mass of SNe II from the PTF sample. The vertical blue lines display the host masses of PTF SNe II with flash features\cite{Khazov2016}. These hosts probe a wide range from $10^8$\,M$_\odot$ to $10^{11}$\,M$_\odot$. Hosts similar to that of SN 2019hgp (shown in red) are fairly common among SNe II with flash features.

The GTC SN spectrum from 29 July 2019 (Supplementary Table~\ref{spec_tab}) shows narrow emission lines from the underlying HII regions. We measure the following line fluxes for H$\alpha$, H$\beta$, [OIII]$\lambda4960$, [OIII]$\lambda5007$, and [NII]$\lambda6585$ of $16.5\pm1.1$, $5.3\pm1.4$, $3.0\pm1.2$, $8.4\pm1.6$ and $3.0\pm0.9$ $\times10^{-18}\,{\rm erg\,cm}^{-2}\,{\rm s}^{-1}$. 
Due to the lack of accurate photometry of the transient at the time of the spectroscopic observation, each measurement can be off by a numerical factor. However, flux ratios of lines close in wavelength space are unaffected by this uncertainty and by the uncertain dust extinction at the explosion site. Therefore, we can estimate the metallicity at the explosion site using the O3N2 indicator with the calibration reported in \cite{Marino2013}. The oxygen abundance of $12+\log({\rm O/H})=8.29^{+0.04}_{-0.05}$ translates to a low metallicity of Z=$0.4\pm0.04$ Z$_{\odot}$ (assuming a solar oxygen abundance of 8.69\cite{Asplund2009}).

Overall, the properties of this galaxy are similar to those of the hosts of other RETs\cite{Ho2020}, as well as those of the host galaxies of hydrogen-poor Type I superluminous SNe (SLSNe-I) and long-duration Gamma-Ray Bursts (GRBs) at $z\sim0.3$\cite{Lunnan2014,Leloudas2015,Perley2016,Schulze2018,Schulze2020}.

\subsection{9. Circumstellar emission in other SN Types:}

We compare our 27.4\,d spectrum of SN 2019hgp with representative spectra of other types of interacting SNe of Type Ibn and IIn in Extended Data Fig.~7. The spectrum is quite similar to those of SNe Ibn, in both the non-thermal continuum shape and some of the features, but it remarkably lacks the strong He I emission lines which are the spectroscopic hallmark of 
Type Ibn SNe. The blue quasi-continuum seen below $5500$\AA\ likely arises from emission from multiple Fe II transitions (as seen for other events\cite{Kiewe2012,Karamehmetoglu2019}). 

%Recently, two additional members of the new class of SNe Icn have been reported\cite{Perley2021a,Pastorello2021}. Supplementary Fig.~\ref{} shows a comparison of example spectra of these objects with our observations of SN 2019hgp, supporting this association. 

\subsection{10. Data on rise times of various transient source classes: }

Fig.~\ref{fig:risetime} plots the peak red-light ($r$ or $R$-band) absolute magnitudes vs. the transient rise time from estimated explosion to peak. As these sources are all nearby, time-dilation corrections are negligible and have not been applied. High-cadence wide-field surveys are especially well suited to determine these parameters, and in particular to accurately estimate the time of explosion, and most data plotted come from such surveys. In particular, data have been extracted from the following sources. Data for SN 2019hgp are from this work. Rise time data for Type II SNe are based on samples from PTF\cite{Rubin2016} and ZTF\cite{Bruch2020}. Data for SNe Ia are from the ZTF sample: peak magnitudes\cite{Yao2019} and rise times\cite{Miller2020}.  
Data for SNe Ic are taken from the PTF samples of normal\cite{Barbarino2020} and broad-line\cite{Taddia2019} events. Additional events with well-determined parameters include SN 2002ap\cite{Gal-Yam2002,Mazzali2002}, SN 1998bw\cite{Galama1998}, SN 2006aj\cite{Campana2006,Bianco2014}, SN 1994I\cite{Richmond1996,Sauer2006}, and PTF12gzk\cite{Ben-Ami2012b}. Data for SNe Ibn are from the high-cadence ZTF survey (Kool et al., in preparation). Unfortunately no similar survey sample data exist yet for SNe Ib, and we compiled data for the well-observed events iPTF13bvn\cite{Cao2013}, SN 1999ex\cite{Stritzinger2002}, SN 2008D\cite{Mazzali2008,Soderberg2008}, and SN 2009jf\cite{Valenti2011}. The locations of a sample of Pan-Starrs 1 Rapidly-Evolving Transients (RETs)\cite{Drout2014} that lack spectroscopic classification are marked with open black markers; additional well-observed RETs included are KSN15K\cite{Rest2018}, iPTF16asu\cite{Whitesides2017}, AT2018cow\cite{Prentice2018,Perley2019b,Margutti2019} and SN 2018gep\cite{Ho2019}.  
RETs iPTF16asu and SN 2018gep show SN-Ic-like spectra during their evolution, while the rapidly-rising event iPTF14gqr, standing out from the rest of the PTF SN Ic sample (green) was suggested to arise from an ultra-stripped progenitor\cite{De2018}; SN 2018dge is a similar event with Type Ib spectral features\cite{Yao2020}. 
The single peculiar Iax event within the ZTF SN Ia sample is marked by an open blue symbol. 

\subsection{11. Modelling the observations:}

We first summarize the main observational properties that any physical models of this event need to confront.

$\bullet$ The bolometric light curve (Fig.~\ref{fig:LC_Bol}) rapidly rises (within $<1.5$\,d) to a luminous peak (L$=3.4\times10^{43}$\,erg\,s$^{-1}$). The timescales of rise and decline are short compared to typical Type I SNe (Fig.~\ref{fig:risetime}; Fig.~\ref{fig:LC_Bol}; Extended Data Fig.~5).

$\bullet$ Our observations are well fit by BB SEDs till day 12 (Extended Data Fig.~4), with 
BB temperatures that rapidly cool from an initially hot peak (T$=30$kK assuming negligible host extinction; possibly as hot as $100$kK for the maximal allowed extinction values of E$_{\rm B-V}=0.15$\,mag; Extended Data Fig.~9).

$\bullet$ The BB radius evolution suggests a free (ballistic) expansion at v=9900 km/s till day 10 (Fig.~\ref{fig:LC_Bol}, inset).

$\bullet$ The event occurred within an expanding wind with a composition dominated by C/O/Ne (Fig.~\ref{fig:spec}), suggesting the  progenitor envelope is also free of hydrogen, and depleted of helium. The wind expansion velocity is high v$_{\rm wind}\sim2000$\,km\,s$^{-1}$ (Fig.~\ref{fig:blue_edge}).

$\bullet$ An ejecta component expanding at typical SN photospheric velocities (v$\sim10,000$\,km\,s$^{-1}$) appears around 19\,d after explosion (Extended Data Fig.~3) and reveals absorption lines of common intermediate-mass elements (O, Na, Mg, Ca), as well as absorption by iron and quite likely neon (Extended Data Fig.~11).

$\bullet$ Observations starting around 15\,d show a UV excess above the best-fit black body (Extended Data Fig.~4); spectral comparison to other types of interacting SNe (Extended Data Fig.~7) shows a blue continuum excess starting at approximately the same time. 

Next, we consider several classes of models and confront them with our observations.

\paragraph{Radioactive $^{56}$Ni:}

Fig.~\ref{fig:LC_Bol} shows the best-fit $^{56}$Ni model found using \textit{Tigerfit}\cite{Tigerfit}. This model requires the entire ejecta to be composed of $^{56}$Ni (with a mass of $0.4$\,M$_{\odot}$); this is driven by the requirement of high Ni mass to attempt to explain the luminous peak while the total ejecta mass is constrained by the rapid rise and decline (short diffusion time) to be low. The resulting solution of having a pure Ni ejecta still misses the peak, has to assume a very low $\gamma$-ray trapping, and is in strong conflict with our spectroscopic observations that are not dominated by iron-group elements at any phase. We thus find that our early photometric data cannot be explained by models based on energy release from freshly synthesized radioactive $^{56}$Ni\cite{Arnett1982}, as is commonly assumed for hydrogen-deficient (Type I) supernovae\cite{Drout2011,Taddia2018,Prentice2019}. A comparison of our bolometric light curve to that of a relatively rapidly-evolving SN Ic (SN 2007gr\cite{Sharon+Kushnir2020}) shows that even scaling this light curve down arbitrarily, no section of our light curve is consistent with the Ni decline slope, indicating that any radioactive contribution is sub-dominant at all observed phases. Models of SN 2007gr\cite{Mazzali2010} suggest the total C/O-dominated ejecta mass of that object is $<2$\,M$_{\odot}$. The comparatively rapid evolution of SN 2019hgp therefore suggests that for any model assuming a centrally-located energy source, the total mass of the ejecta (also dominated by C/O in our case, with similar expansion velocities; Extended Data Fig.~3) would be smaller than this value, in order for the diffusion time to be shorter.     

\paragraph{Pure CSM interaction:}

Fig.~\ref{fig:LC_Bol} shows that a simple CSM interaction model\cite{Chatzopoulos2012} describes the Bolometric light curve well throughout its evolution, and the derived best-fit parameters (progenitor radius of R$_*=4.1\times10^{11}$\,cm, ejecta mass of M$_{\rm ej}= 1.2$\,M$_{\odot}$, opacity of $\kappa = 0.04$\,cm$^2$\,g$^{-1}$, CSM mass M$_{\rm CSM}=0.2$\,M$_{\odot}$, and a mass loss rate $\dot{\rm M}=0.004$\,M$_{\odot}$\,y$^{-1}$ expanding at a velocity of 
v$_{wind}=1900$\,km\,s$^{-1}$) are remarkably consistent with the values we estimate directly from the data. We note that these models are simple and include several assumptions, most notably that the reverse and forward shock heating are both centrally located, and  terminate when the SN ejecta have been swept up by the reverse shock, and the forward shock breaks out of the CSM. This simplified assumption of centrally located shocks can lead to an overestimated diffusion timescale and underestimated CSM and ejecta masses\cite{Chatzopoulos2012}.
Yet, this interpretation faces two major difficulties. The first is the observed spectroscopic evolution of SN 2019hgp. While the initial spectra (Extended Data Fig.~1-2) show narrow lines superposed on a blue continuum, as seen in other interacting transients of types IIn and Ibn (Extended Data Fig.~7), starting at day 19 (Extended Data Fig.~3), our spectra show broad absorption features with high expansion velocities (v$=10,000$\,km\,s$^{-1}$) which suggest we are seeing the supernova ejecta directly, rather than emission from shocked CSM. This requires a different energy source for the emission at later phases. A second conundrum with the pure CSM model is that during the initial 10 days after explosion, the emitting region smoothly expands with a constant velocity (Fig.~\ref{fig:LC_Bol}; inset). This behaviour cannot be accommodated in a simple spherical CSM interaction model, and would require a non-spherical geometry\cite{Soumagnac2019b}. Interestingly, non-spherical CSM geometry has been observed around WR stars\cite{Smith2011b}. We thus conclude that CSM interaction is likely important in this event, but a simple, spherical CSM interaction model that assigns the entire emitted energy to interaction is inconsistent with the data. An important caveat for CSM models is that the ejecta mass estimate includes only the ejecta that take part in the interaction (typically the fastest, external layers) while the mass of more slowly-moving material is unconstrained. With an additional, large unobserved mass component, the total ejecta mass may become consistent with a neutron star (rather than a black hole) remnant.        

\paragraph{Shock cooling within a CSM nebula:} 

In analogy to Type II SNe, one may consider a model where the ejecta are heated by the explosion shock and slowly radiate this energy (the shock-cooling emission) over an extended period of time. In Type II SNe this model is commonly considered, and the spectroscopic behavior seen, with a blue continuum initially (with superposed emission lines in objects embedded in CSM\cite{Gal-Yam2014,Khazov2016,Yaron2017,Bruch2020}) evolving to a photospheric spectrum with broad absorption features\cite{agy17}, is broadly similar to what we observe here. However, as can be seen in Supplementary Fig.~\ref{fig:RW}, in order to reach the peak bolometric luminosity we measure (L$=3.44\times10^{43}$\,erg\,s$^{-1}$), a supergiant progenitor (with R$_{*}>10^{12}$\,cm) is needed, for any reasonable explosion energy, in contrast to the compact progenitor indicated by our spectroscopic data (Fig.~\ref{fig:blue_edge}). The modest expansion velocity we measure (v$=10,000$\,km\,s$^{-1}$) for our low-mass ejecta (M$<2\,$M$_{\odot}$) in fact suggests a low kinetic energy, making the radius constraint stricter. We therefore conclude that a standard shock-cooling model within a CSM distribution does not fit our observations.        

\paragraph{Shock breakout in a wind:}

The first electromagnetic signature arriving to a distant observer from an exploding star is a flare of radiation emitted when the explosion shock breaks out from the stellar surface (the shock-breakout flare\cite{Waxman+Katz2017}). For a compact star as we consider here, the shock breakout emission peaks at high energy and would be too weak to be observed in visible light by ZTF\cite{Ganot2016}. However, if the star is embedded in a thick wind, as may be the case here, the breakout occurs in the wind, at a radius much larger than that of the progenitor. Such wind-breakout flares are much longer and more luminous than stellar breakouts, and could peak in the near UV\cite{Ofek2010}, making this a plausible model for SN 2019hgp. In fact, SN 2009uj, an interacting transient suggested to result from a wind breakout\cite{Ofek2010}, has a UV rise timescale similar to that of SN 2019hgp (7\,d and 4\,d, respectively) and an almost identical $r-$band decline slope. To test this idea, we estimate the expected BB temperature during such a flare. This could be done by applying eq. 7 from\cite{Ofek2010}, $T=9.1\times10^4$\,$\kappa_{0.34}^{-1/4}$\,t$_{7}^{-1/4}$K. Here $\kappa_{0.34}$ is the opacity in cm$^2$\,g$^{-1}$, and t$_7$ is the time since explosion in units of 7\,days. This time requires some attention, as the explosion time we used so far was actually the time of first light. The difference is the time it takes the explosion shock to propagate within the star, which is negligible ($<0.005$\,d for a $10,000$\,km\,s$^{-1}$ shock propagating in a compact star with R$_*\sim4\times10^{11}$\,cm) compared to our estimated uncertainties ($0.2$\,d). However, this propagation time is not negligible for the larger wind radii we consider here. If we adopt the intercept of the BB radius evolution at the time of first light (Fig.~\ref{fig:LC_Bol}, inset) as the wind breakout radius ($\sim2\times10^{14}$\,cm), and add the propagation time for a $10,000$\,km\,s$^{-1}$ shock (2.3\,d) to the time from first light till our first SED was obtained (1.5\,d, Extended Data Fig.~4), we find using the equation above a predicted temperature range T$=1.2\times10^5-1.8\times10^5$\,K for opacity values $\kappa=0.2$\,cm$^2$\,g$^{-1}$ and $\kappa=0.04$\,cm$^2$\,g$^{-1}$, respectively, which bracket the range of highly ionized He/C/O mixtures\cite{Rabinak2011}.  
This estimate is consistent with the upper range of the allowed temperature during this epoch assuming the extinction in the host is not negligible (Extended Data Fig.~9). We can conclude that our observations are not in conflict with a wind breakout powering the peak of the emission seen. However, an additional mechanism, possible interaction, is likely required to power the UV excess and blue spectral continuum seen later during the evolution of this object.

\paragraph{Model summary:} 

Having studied several possible models for our observations, it appears that no single simple idea can explain all the observations. Some models (e.g., $^{56}$Ni radioactivity and shock-cooling emission), are unlikely to significantly contribute. In fact, the failure of $^{56}$Ni models can be taken as a defining feature of RETs, such as SN 2019hgp, and its Ibn and Icn cousins. On the other hand, CSM interaction is very likely to play a part in explaining the observations. While a simple spherical interaction model is problematic, interaction is likely required to explain the late-time UV excess and blue spectral continuum, and is expected given the progenitor obviously exploded within a CSM nebula. Solutions to difficulties encountered at late time (the appearance of broad absorption features) could include certain geometries, such as a CSM torus seen from an angle close to the polar direction; in this way the observer sees both the expanding ejecta directly and the interaction emission from the ejecta hitting the inner radius of the torus\cite{Smith2015}. Alternatively the CSM may be clumpy; both options have been discussed before\cite{Ben-Ami2014}. A hybrid model (e.g., a wind breakout followed by an interaction phase) may be an attractive option to explain our rich data set.   

%Models:
%- Wind SB (+ interaction) - ben-ami, clumping
% Where to discuss PPISN? 

\end{methods}

%%%%%%%%%%%%%%%%%%%%%%%%%%%%%%%%%%%%%%%%%%%%%%%%%%%%%%%%%%%%%%%%%%%%%%%%%%%%
%%%%								BIBLIOGRAPHY  2
%%%%%%%%%%%%%%%%%%%%%%%%%%%%%%%%%%%%%%%%%%%%%%%%%%%%%%%%%%%%%%%%%%%%%%%%%%%%
%%%

%ADS custom format: \\bibitem{%2h%Y}\n\bibinfo{author}{%1G}\n\newblock \bibinfo{title}{%T}.\n\newblock \emph{\bibinfo{journal}{%J}} \textbf{\bibinfo{volume}{%V}},\n\bibinfo{pages}{%p} (\bibinfo{year}{%Y}).

%%%%%%%%%%%%%%%%%%%%%%%%%%%%%%%%%%%%%%%%%%%%%%%%%%%%%%%%%%%%%%%%%%%%%%%%%
%								ADDENDUM
%%%%%%%%%%%%%%%%%%%%%%%%%%%%%%%%%%%%%%%%%%%%%%%%%%%%%%%%%%%%%%%%%%%%%%%%%

 \clearpage
 
 \begin{addendum}

\item[Correspondence] Correspondence and requests for materials
should be addressed to Avishay Gal-Yam~(email: avishay.gal-yam@weizmann.ac.il).

\item 
This work is based on observations obtained with the Samuel Oschin 48-inch Telescope and the 60-inch Telescope at Palomar Observatory as part of the Zwicky Transient Facility project. ZTF is supported by the U.S. National Science Foundation (NSF) under grant AST-1440341 and a collaboration including Caltech, IPAC, the Weizmann Institute for Science, the Oskar Klein Center at Stockholm University, the University of Maryland, the University of Washington, Deutsches Elektronen-Synchrotron and Humboldt University, Los Alamos National Laboratories, the TANGO Consortium of Taiwan, the University of Wisconsin at Milwaukee, and Lawrence Berkeley National Laboratories. Operations are conducted by COO, IPAC, and UW. 
This work includes observations made with the Nordic Optical Telescope (NOT), owned in collaboration by the University of Turku, and Aarhus University, and operated jointly by Aarhus University, the University of Turku, and the University of Oslo (representing Denmark, Finland, and Norway, respectively), the University of Iceland, and Stockholm University, at the Observatorio del Roque de los Muchachos, La Palma, Spain, of the Instituto de Astrofisica de Canarias. These data were obtained with ALFOSC, which is provided by the Instituto de Astrofisica de Andalucia (IAA) under a joint agreement with the University of Copenhagen and NOT.
This work includes observations made with the GTC telescope, in the Spanish Observatorio del Roque de los Muchachos of the Instituto de Astrofísica de Canarias, under Director's Discretionary Time.
Some of the data presented herein were obtained at the W. M. Keck Observatory, which is operated as a scientific partnership among the California Institute of Technology, the University of California, and the National Aeronautics and Space Administration (NASA); the Observatory was made possible by the generous financial support of the W. M. Keck Foundation. This work includes observations obtained at the international Gemini Observatory, a program of NSF's NOIRLab, which is managed by the Association of Universities for Research in Astronomy (AURA) under a cooperative agreement with the NSF on behalf of the Gemini Observatory partnership: the NSF (United States), National Research Council (Canada), Agencia Nacional de Investigacion y Desarrollo (Chile), Ministerio de Ciencia, Tecnología e Innovacion (Argentina), Ministerio da Ciencia, Tecnologia, Inovacoes e Comunicacoes (Brazil), and Korea Astronomy and Space Science Institute (Republic of Korea). The authors wish to recognize and acknowledge the very significant cultural role and reverence that the summit of Maunakea has always had within the indigenous Hawaiian community.  We are most fortunate to have the opportunity to conduct observations from this mountain.
This work includes observations obtained at the Liverpool Telescope, which is operated on the island of La Palma by Liverpool John Moores University in the Spanish Observatorio del Roque de los Muchachos of the Instituto de Astrofisica de Canarias with financial support from the UK Science and Technology Facilities Council.
Research at Lick Observatory is partially supported by a generous gift from Google.
This work includes observations obtained with the Hobby-Eberly Telescope, which is a joint project of the University of Texas at Austin, the Pennsylvania State University, Ludwig-Maximilians-Universit\"at M\"unchen, and Georg-August-Universit\"at G\"ottingen.
These results made use of the Lowell Discovery Telescope (LDT) at Lowell Observatory. Lowell is a private, nonprofit institution dedicated to astrophysical research and public appreciation of astronomy and operates the LDT in partnership with Boston University, the University of Maryland, the University of Toledo, Northern Arizona University, and Yale University.
This work benefited from the OPTICON telescope access program (\url{https://www.astro-opticon.org/index.html}), funded from the European Union's Horizon 2020 research and innovation programme under grant agreement 730890. We made use of IRAF, which is distributed by the NSF NOIR Lab.

A.G.-Y. is supported by the EU via ERC grant No. 725161, the ISF GW excellence center, an IMOS space infrastructure grant, BSF/Transformative and GIF grants, as well as by the Benoziyo Endowment Fund for the Advancement of Science, the Deloro Institute for Advanced Research in Space and Optics, The Veronika A. Rabl Physics Discretionary Fund, Minerva, Yeda-Sela and the Schwartz/Reisman Collaborative Science Program; A.G.-Y. is the incumbent of the The Arlyn Imberman Professorial Chair.
M.M.K. acknowledges generous support from the David and Lucille Packard Foundation; the GROWTH project was funded by the NSF under grant AST-1545949.
E.C.K., J.S, and S.S. acknowledge support from the G.R.E.A.T. research environment funded by {\em Vetenskapsr\aa det}, the Swedish Research Council, under project number 2016-06012; E.C.K. also received support from The Wenner-Gren Foundations.
O.K.C.'s participation in ZTF was made available by the K.A.W. Foundation.
G.L. is supported by a research grant (19054) from VILLUM FONDEN.
J.C.W. and B.P.T. are supported by NSF grant AST-1813825. A.V.F.'s supernova group at U.C. Berkeley is supported by the TABASGo Foundation, the Christopher R. Redlich Fund, and the Miller Institute for Basic Research in Science (A.V.F. is a Senior Miller Fellow).

\item[Author Contributions] 
A.G.-Y. initiated the project, planned the observations, conducted spectroscopic and physical analysis, and wrote the manuscript.
R.J.B. identified the transient, initiated follow-up observations, conducted photometric analysis, and contributed to the WIS infant SN program.
S.S. contributed to follow-up design and execution, conducted multiwavelength and host-galaxy analysis, and contributed to the manuscript.
Y.Y. contributed to follow-up design and execution, reduced the Gemini spectra, and contributed to the manuscript.
D.A.P. conducted follow-up observations with the LT and contributed to the manuscript.
I.I. conducted photometric and spectroscopic analysis and contributed to physical interpretation and manuscript writing.
J.S. helped to plan and develop the manuscript and provided NOT data. 
E.C.K. analyzed the bolometric light curve and provided physical modelling. 
M.T.S. contributed to photometric analysis.
O.Y. conducted spectroscopic modelling. N.L.S. conducted a prediscovery variability search and contributed to the manuscript.
E.Z. contributed to spectroscopic analysis.
C.B. Provided the sample of SNe~Ic from PTF and reduced the NOT spectroscopy.
S.R.K. is the ZTF PI. 
S.R.K., M.M.K., C.F., and L.Y. provided Palomar and Keck data.
K.D. and Y.Y. reduced the Palomar and Keck data. 
E.O.O. and C.F. contributed to physical interpretation.
A.V.F., W.Z., T.G.B., R.J.F., J.B., and M.S. contributed Lick and Keck data; A.V.F. also contributed to the manuscipt.
C.M.C. contributed LT data. 
A.L.C.-L., D.G.-A., and A.M.-B. provided GTC observations.
S.F. and T.H. provided LDT observations.
J.C.W., B.P.T., and J.V. planned, obtained, and reduced the HET observations. 
G.L. reduced the GTC spectra and contributed to the manuscript.
M.J.G., D.A.D., A.J.D., R.D., E.C.B., B.R., D.L.S., I.A., Y.S., R.R., and J.v.R are ZTF builders.
N.K. contributed to the spectroscopic analysis.
Many authors provided comments on the manuscript, and all authors have approved it.

 \item[Competing Interests] The authors declare that they have no
competing financial interests.

\item[Data Availability Statement]
The photometry of SN 2019hgp is available in Supplementary Table~1, and all the observations (photometry and  spectra) are available from WISeREP\cite{yg12,WISeREP}. Matlab scripts that generate most of the plots within this paper are available from the corresponding author upon request. Opticon observations were obtained under Program ID OPT/2019A/024, PI Gal-Yam.

\item[Code Availability Statement]
Relevant software sources have been provided in the text, web locations provided as references, and are publicly available.

\end{addendum}

\setcounter{enumiv}{\value{firstbib}} 

%%%%%%%%%%%%%%%%%%%%%%%%%%%%%%%%%%%%%%%%%%%%%%%%%%%%%%%%%%%%%%%%%%%%%%%%%
%								SUPPLEMENTARY FIGURES
%%%%%%%%%%%%%%%%%%%%%%%%%%%%%%%%%%%%%%%%%%%%%%%%%%%%%%%%%%%%%%%%%%%%%%%%%

\clearpage

\begin{addendum}

 \item[Extended Data Items]
 
 \clearpage
 
 \begin{figure}
\centering
\vspace*{-3cm}
\hspace*{-0.5cm}\includegraphics[width=16cm]{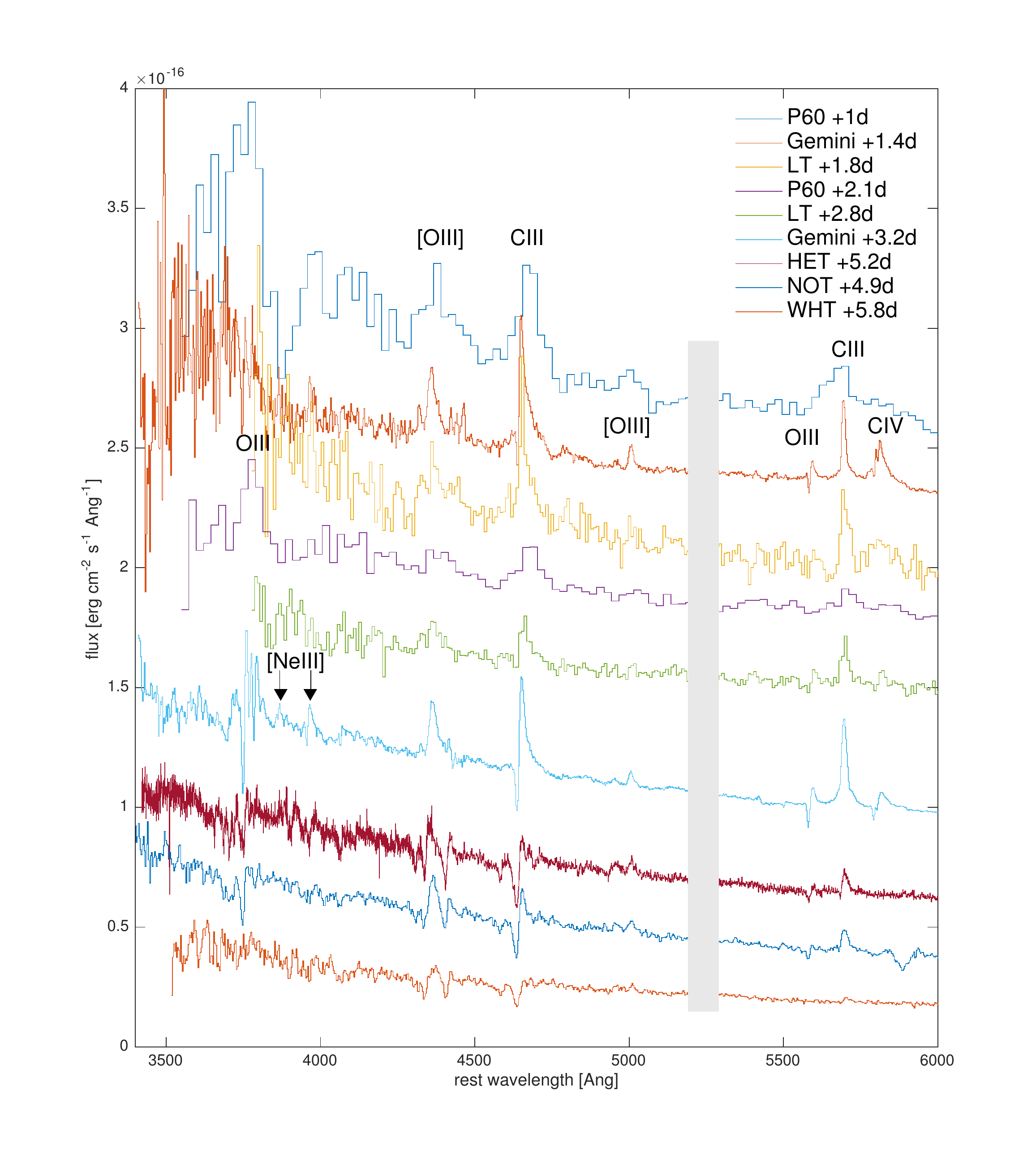}
\vspace*{-2cm}
% commented our caption in order to suppress the figure numbering in these data item figures
% need to also give up on labels since the pointers do not go anywhere

%\caption{
~\\
{\bf Extended Data Figure 1~~~~}
The spectroscopic series obtained during the initial hot phase (1-5.6\,d after explosion) shows strong emission lines of highly ionized carbon, oxygen and neon (see Fig.~\ref{fig:spec} for detailed line identification), that weaken with time. Pure emission lines evolve to P Cygni profiles, and then to absorption-dominated profiles. Major emission features are marked; the spectral area around 5250 \AA\ in restframe is impacted by imperfect subtraction of the strong atmospheric 5577 \AA\ sky line (grey shade). Five additional P60 and LT spectra with lower S/N and spectral resolution obtained during this period are omitted for clarity.  
%\label{fig_spec_early}
%}
\end{figure}

\clearpage

\begin{figure}
\centering
\vspace*{-3cm}
\hspace*{-0.5cm}\includegraphics[width=16cm]{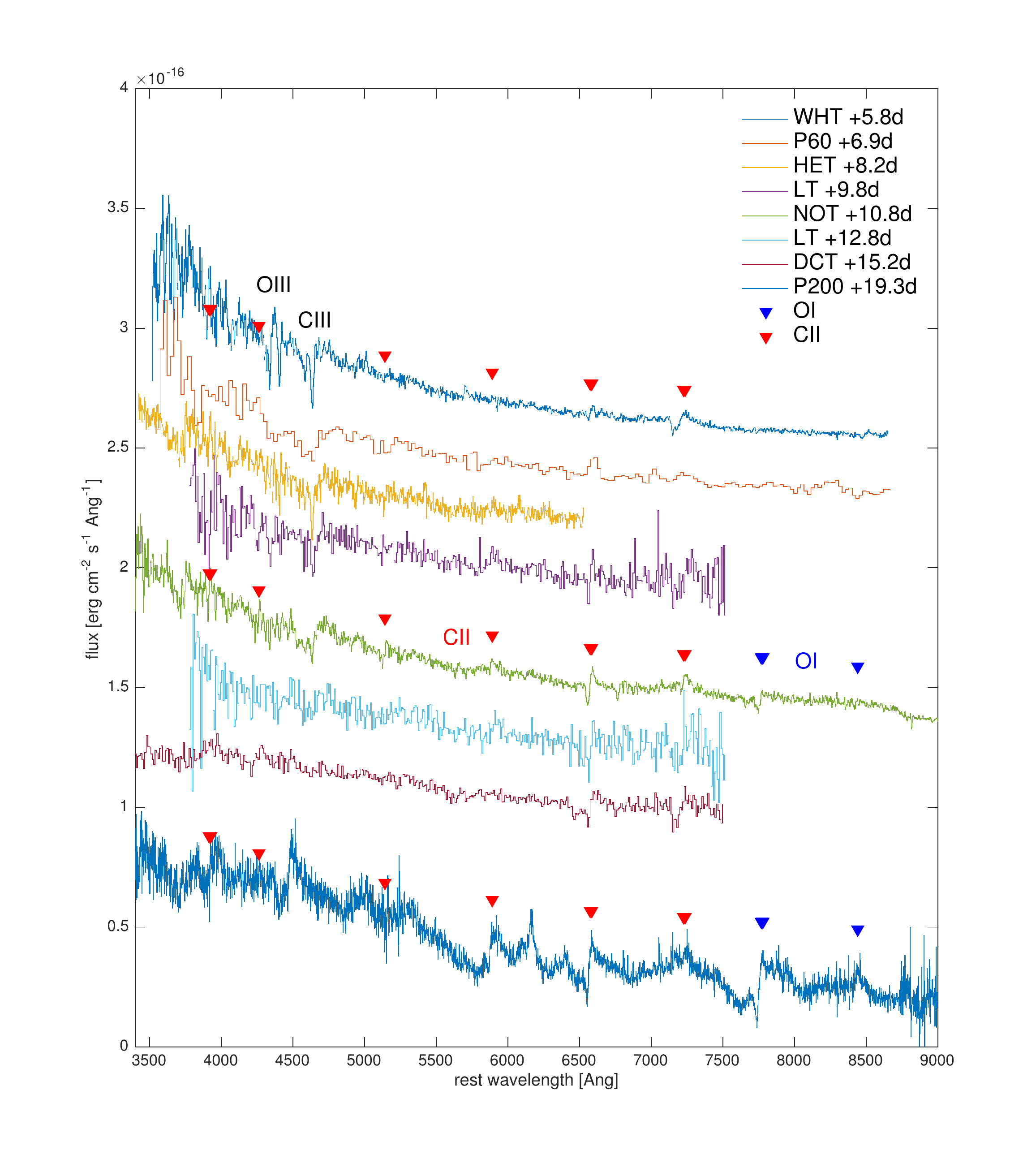}
\vspace*{-1cm}
%\caption{
~\\
{\bf Extended Data Figure 2~~~~}
The spectroscopic series obtained during the intermediate phase (5-19\,d after explosion) follows the weakening and disappearance of the CIII and OIII absorption features seen earlier, and the emergence of a set of low-ionization emission lines, initially of CII (red) and later OI (blue; $50\%$ intensity lines extracted as in \cite{agy18}). Higher resolution spectra resolve the broad features in the blue into multiple narrow components better described by CIII and OIII at zero velocity than by OII blends sometimes seen in hot early phases of stripped SNe, including Type I SLSNe\cite{agy17,agy19}. By day 19 (bottom) broad features appear and the spectrum shows a marked blue excess. Seven additional spectra omitted for clarity.  
%\label{fig_spec_mid}
%}
\end{figure}

\clearpage

\begin{figure}
\centering
\vspace*{-3cm}
\hspace*{-0.5cm}\includegraphics[width=14cm]{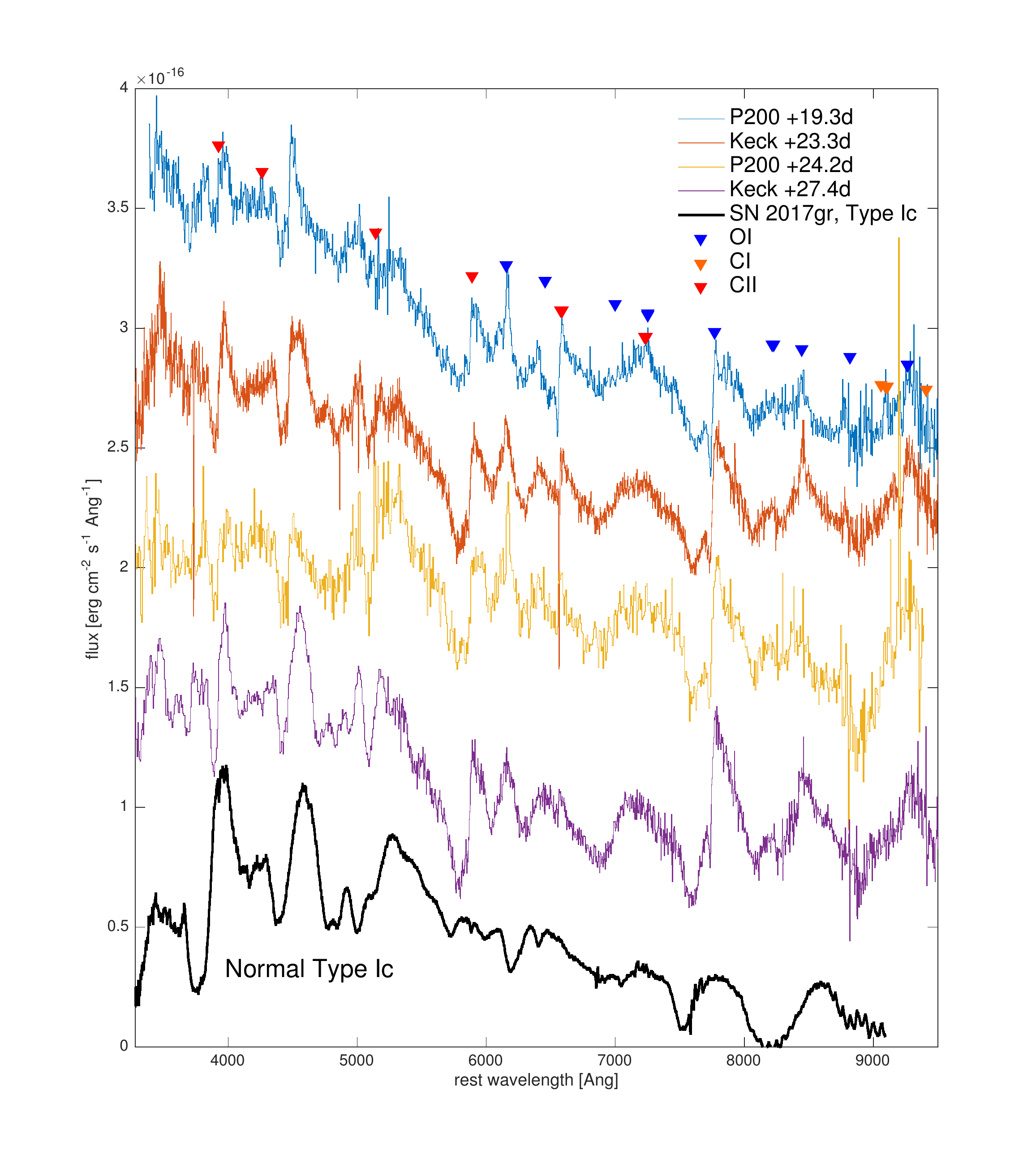}
\vspace*{-0cm}
%\caption{
~\\
{\bf Extended Data Figure 3~~~~}
The spectroscopic series obtained during the late phase (19-27\,d after explosion) 
evolves as features of heavier elements
(e.g., Mg) begin to emerge, while broad
absorption features develop. Initially, strong features (such as OI $7774$\AA\ and CII $6580$\AA) present both a narrow ($\sim2000$\,km\,s$^{-1}$ blue edge) absorption feature as well as a broader ($\sim6000$\,km\,s$^{-1}$ minimum) component. At 27\,d after explosion, relatively broad absorption features have developed that are reminiscent of spectra of Type Ic SNe, with features from Mg, Ca and Fe appearing in addition to C and O. Excess continuum in the blue is evident, likely arising from the Fe II pseudo-continuum often seen in spectra of interacting SNe (Types IIn and Ibn\cite{agy17}).  
%\label{fig_spec_late}
%}
\end{figure}

\clearpage

\begin{figure}
\centering
\vspace*{-3cm}
\hspace*{-0.5cm}\includegraphics[width=18cm]{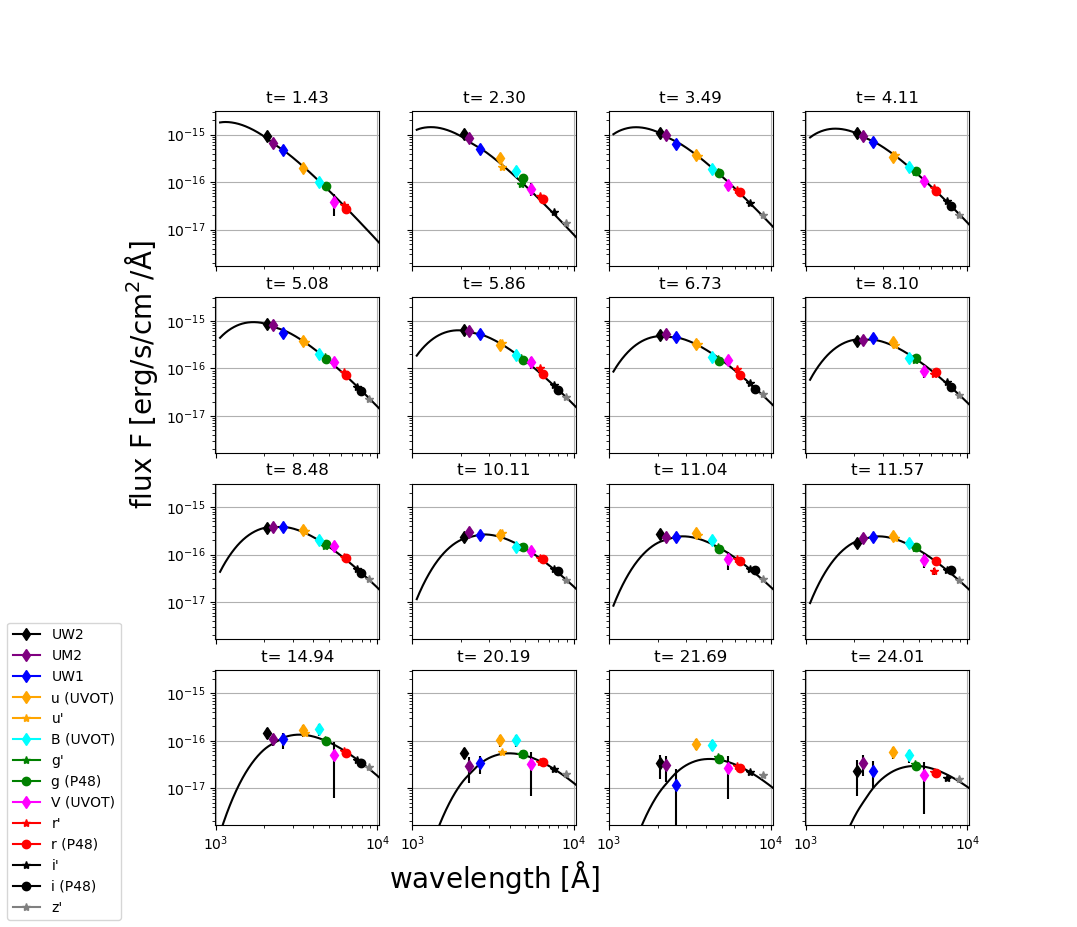}
\vspace*{-0cm}
%\caption{
~\\
{\bf Extended Data Figure 4~~~~}
Black body SED fits calculated using {\tt PhotoFit}\cite{Soumagnac2019a}. Our well-sampled photometry extending from the {\it Swift} ultra-violet (UV) bands to the near-infrared (NIR) z'-band is well-fit by a blackbody curve during the first 12 days after explosion. From day 15 onwards, a clear blue excess develops initially in the UV and extending into the blue part of the optical band from day 21 onward. The derived black-body parameters (radius and temperature) are therefore less reliable from that date. Standard 1$\sigma$ error bars marked.  
%\label{fig:SED}
%}
\end{figure}
 
\clearpage

\begin{figure}
\centering
\vspace*{-3cm}
\hspace*{0.5cm}\includegraphics[width=14cm]{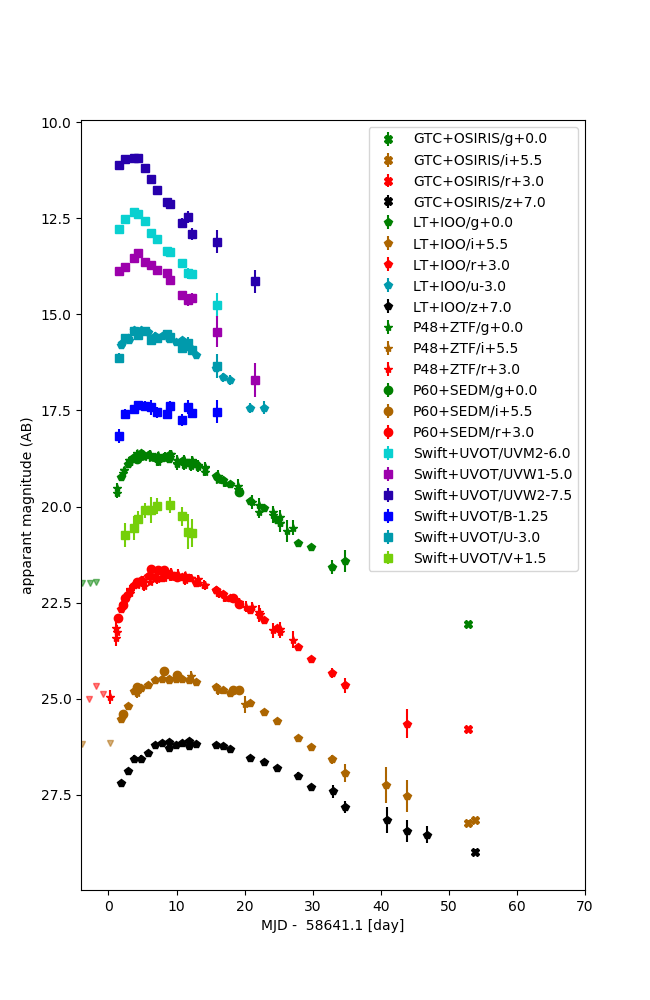}
\vspace*{-1.7cm}
%\caption{
~\\
{\bf Extended Data Figure 5~~~~}
Light curves of SN 2019hgp extending from the UV to the near-IR. Post-peak {\it Swift} $B-$ and $V-$band photometry is inconsistent with data from other sources and likely unreliable. Five outlying P60 points (1 $u$, 2 $g$ and 2 $r$) are inconsistent with the rest of the data to well above their formal errors and have been removed. Standard 1$\sigma$ error bars marked.
%\label{fig:LCs}
%}
\end{figure}
 
\clearpage

\begin{figure}
 \centering
\hspace*{-0.5cm}\includegraphics[width=18cm]{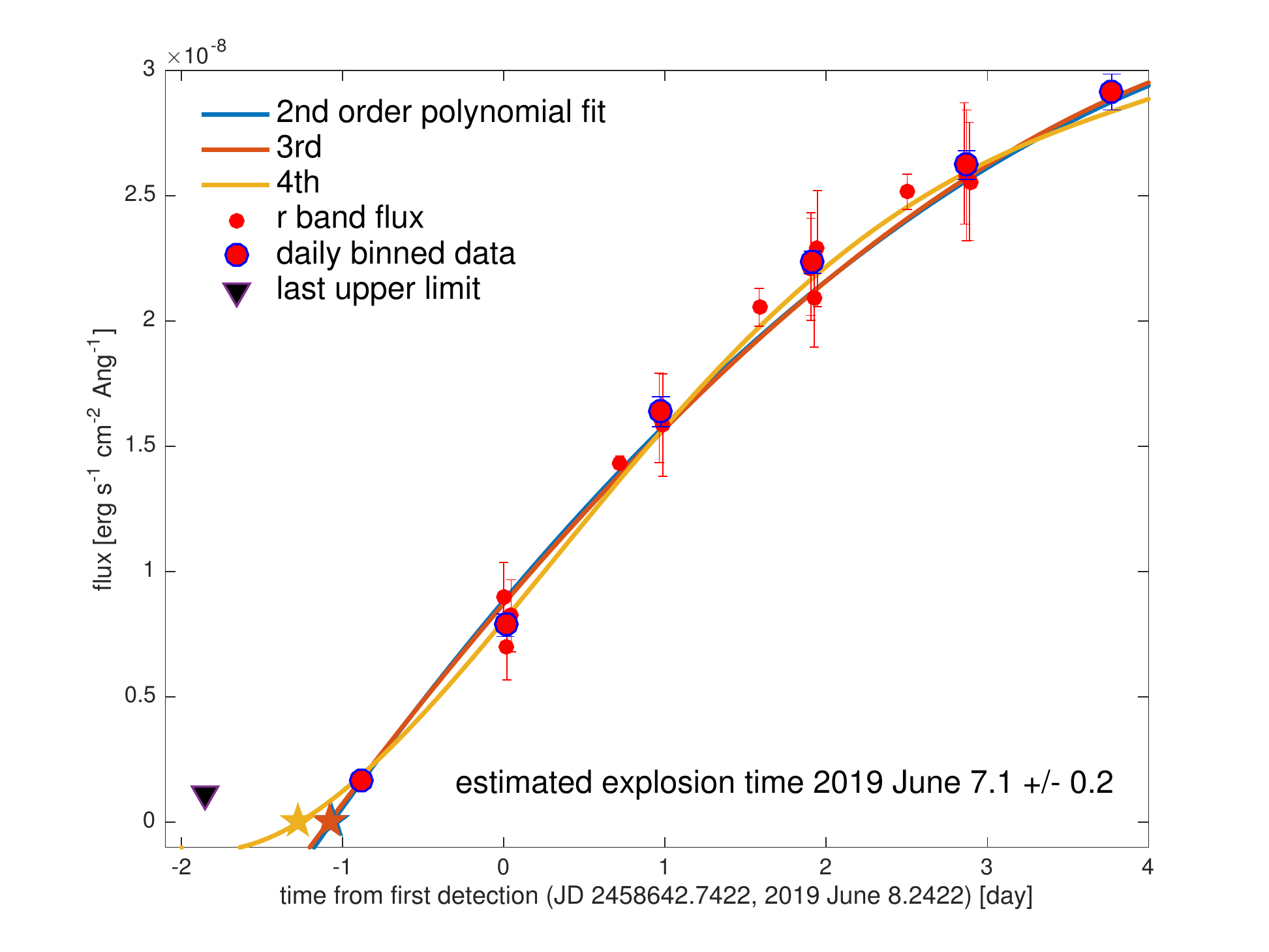}
% commented our caption in order to suppress the figure numbering in these data item figures
% need to also give up on labels since the pointers do not go anywhere

%\caption{
~\\
{\bf Extended Data Figure 6~~~~} Low-order polynomial fits to the early $r$-band photometry indicate the explosion occurred on 2019 June $7.1\pm0.2$\,d. While a linear fit does not provide a good description of the data, low-order (degrees 2-4) polynomials fit the data well and converge on an estimated explosion time occurring $\sim1$\,d prior to discovery (stars denote extrapolated times of zero flux). Stacked pre-discovery data recover a detection during the prior night. All times in the paper are reported relative to this fiducial explosion time. The last 5$\sigma$ non-detection is also marked. Standard 1$\sigma$ error bars marked. 
%\label{fig:exptime_r}
%}
\end{figure}

\clearpage

\begin{figure}
\centering
\vspace*{-1cm}
\hspace*{-0.5cm}\includegraphics[width=17cm]{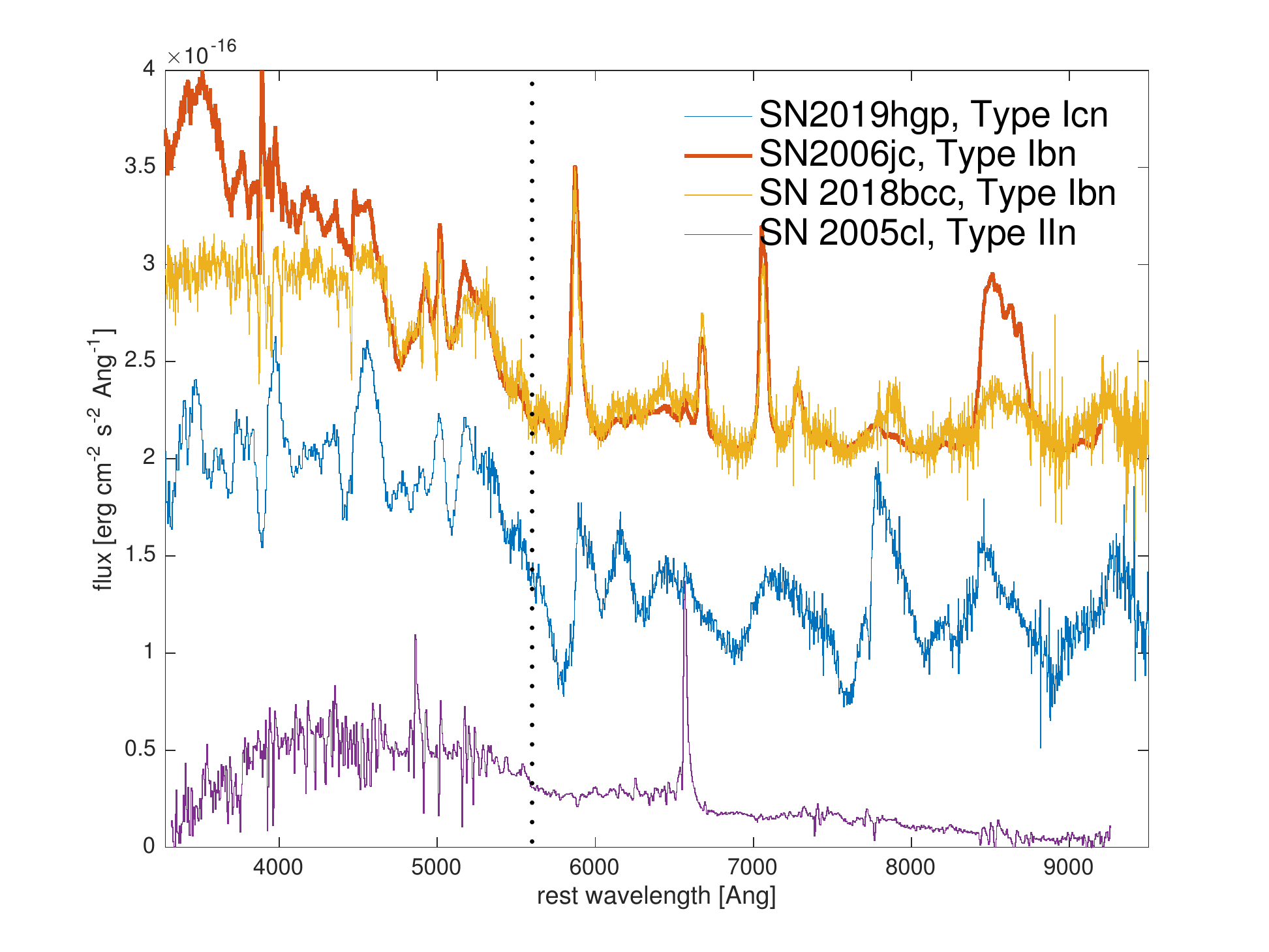}
\vspace*{-1cm}
%\caption{
~\\
{\bf Extended Data Figure 7~~~~}
A comparison of spectra of interacting SNe. Our spectrum of SN 2019hgp is overall quite similar to those of SNe Ibn (SN 2016jc\cite{Foley2007} and SN 2018bcc\cite{Karamehmetoglu2019}), sharing in particular the unusual non-thermal continuum that is flat on the red side, and has a pronounced elevation blueward of $\sim5500$\AA\ (dotted line); this emission likely arises from a quasi-continuum of multiple Fe II emission lines (resolved in some cases, e.g., the Type IIn SN 2005cl\cite{Kiewe2012}, bottom). The hallmark strong He I emission lines common to SNe Ibn ($\lambda\lambda5876,6678,7065,7281$) are absent from the spectrum of SN 2019hgp. Remarkably, this object does, however, show broad absorption features that are missing from spectra of Type Ibn and Type IIn, suggesting that strong shocks are not obscuring our line of sight at 27.4\,d after explosion.
%\label{fig:IbnIcnIIn}
%}
\end{figure}

\clearpage

\begin{figure}
\centering
\vspace*{-1cm}
\hspace*{-0.5cm}\includegraphics[width=14cm]{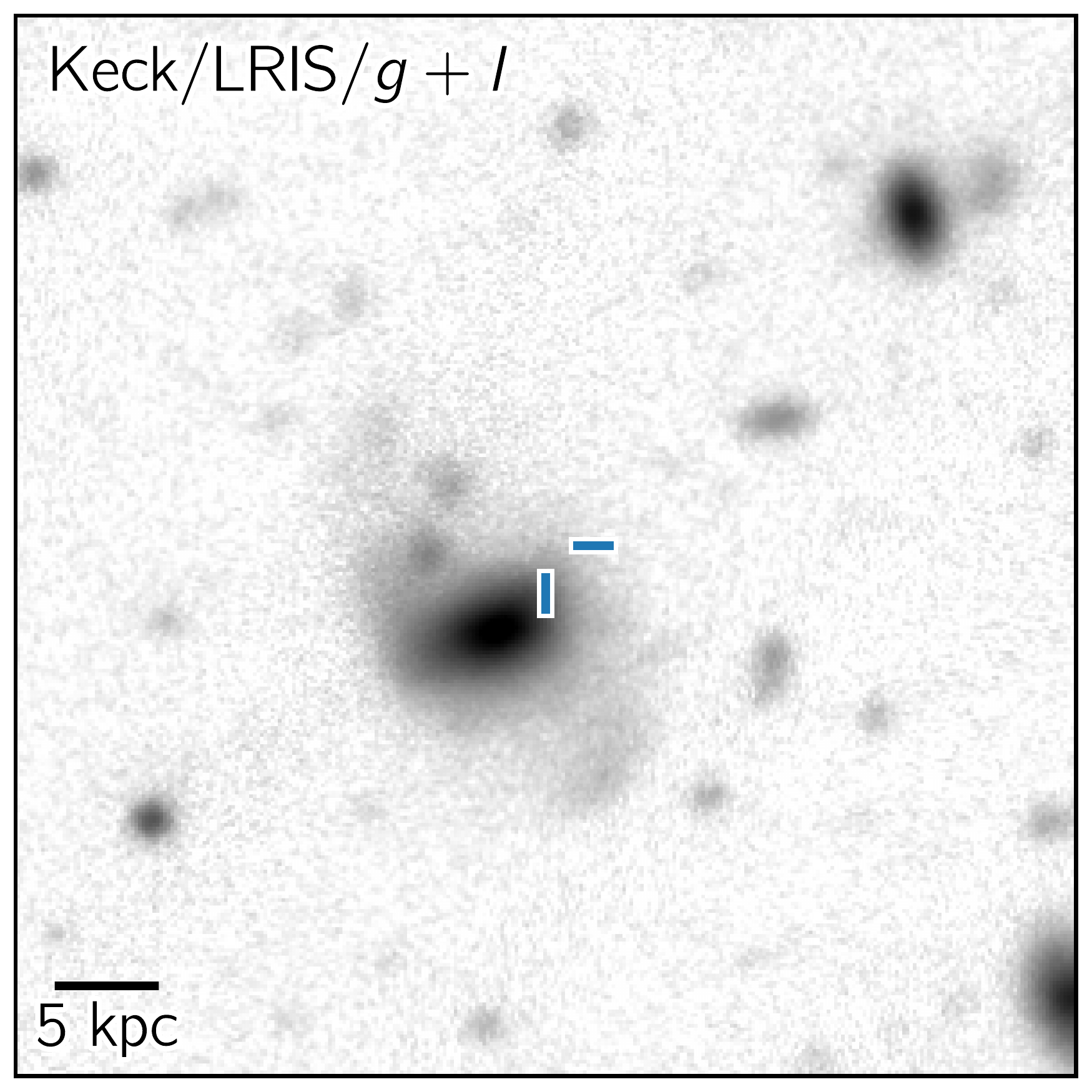}
\vspace*{-0cm}
%\caption{
~\\
{\bf Extended Data Figure 8~~~~}
SN 2019hgp (marked by the crosshair) exploded in the outskirts of its host galaxy at a projected distance of 4.4 kpc (3.54''). The host shows elongated arms of diffuse emission which could suggest a spiral arm or a recent episode of galaxy interaction. In this image East is to the left and North up. The image size is 40" on the side.
%\label{fig:fig_postage_stamp}
%}
\end{figure}

\clearpage

\begin{figure}
\centering
\vspace*{-3cm}
\hspace*{-0.5cm}\includegraphics[width=17.5cm]{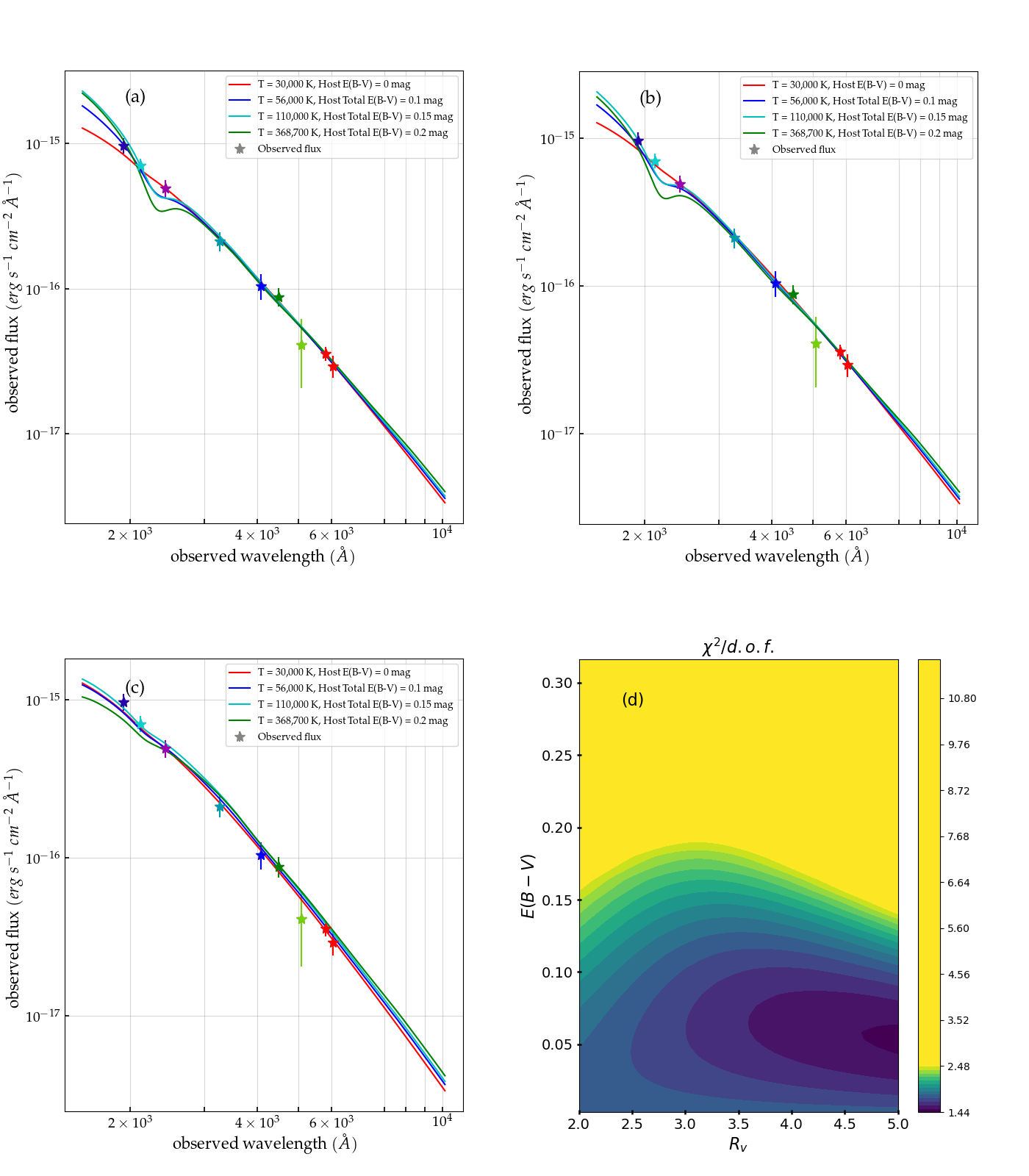}
\vspace*{-1.7cm}
%\caption{
~\\
{\bf Extended Data Figure 9~~~~}
Extinction fits to our first-epoch SED (+1.5\,d) using MW (a), LMC (b) and SMC (c) extinction laws. A fit with negligible host extinction (red) fits the data well. Values of extinction, extending up to E$_{\rm B-V}=0.15$\,mag (requiring BB temperatures of $\sim100$\,kK) are allowed; higher extinction is ruled out regardless of extinction law parameters (MW (d) law shown, SMC and LMC are similar). $\chi^2$ minimization is done using epochs well fit by BB curves ($<15$\,d). Standard 1$\sigma$ error bars marked.
%\label{fig:extinction}
%}
\end{figure}

\clearpage

\begin{figure}
\centering
\vspace*{-3cm}
\hspace*{-0.5cm}\includegraphics[width=14cm]{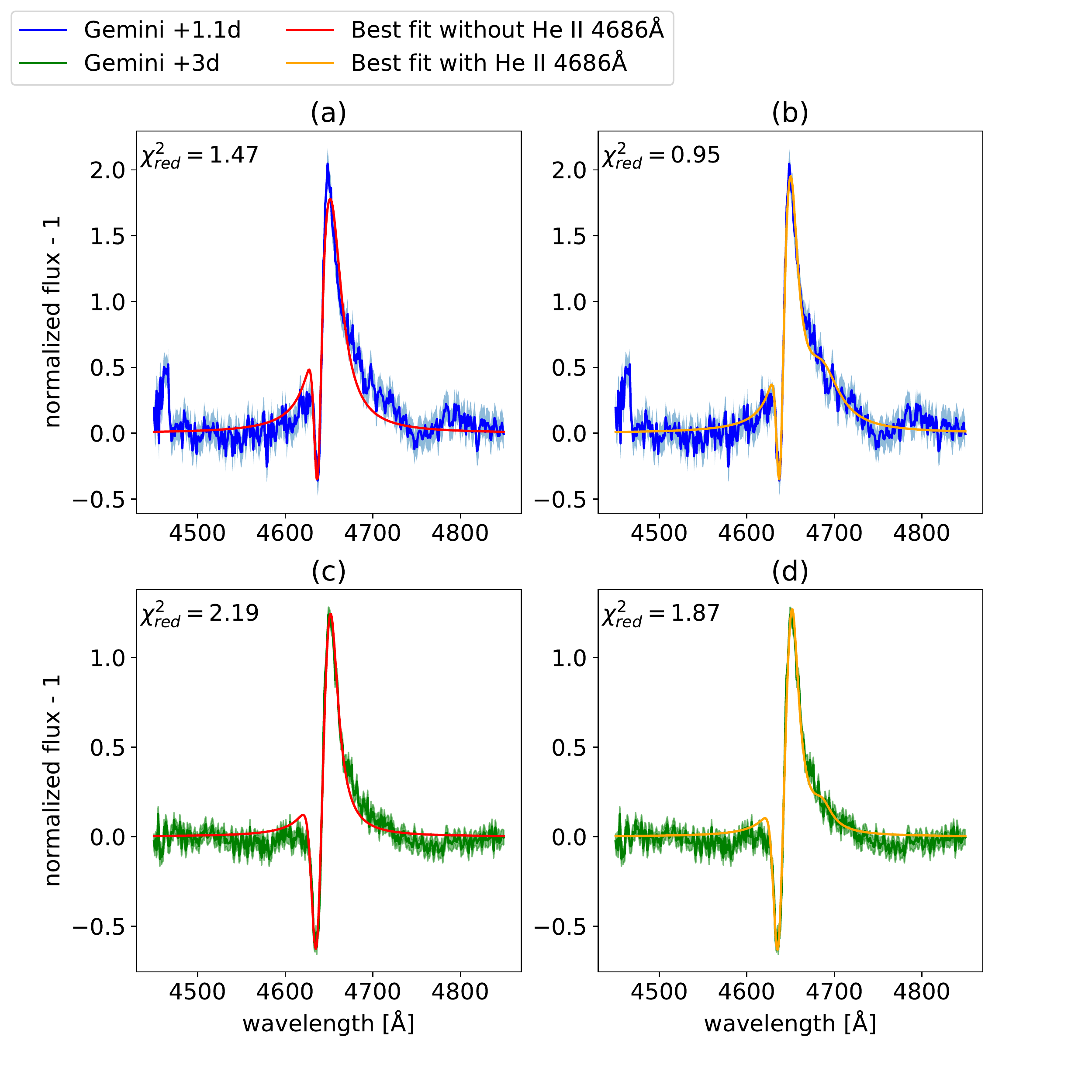}
\vspace*{-0cm}
%\caption{
~\\
{\bf Extended Data Figure 10~~~~}
Modelling of the emission complex around $4660$\AA\ during the first two Gemini epochs (1 and 3 days after explosion, panels (a),(b) and (c),(d), respectively). We fit a combination of a Lorentzian emission component of CIII $\lambda4650$\AA\ along with a blueshifted Gaussian absorption component. Including an additional Lorentzian emission from He II $\lambda$4686\AA\ (panels (b),(d)) is preferred by the data (in the $\chi^2$ sense) even though this feature does not appear as a distinct emission peak. We conclude that the presence of He II in these spectra cannot be ruled out.  
%\label{fig:HeIIchi2}
%}
\end{figure}

\clearpage

\begin{figure}
\centering
\vspace*{-3cm}
\hspace*{-0.5cm}\includegraphics[width=17cm]{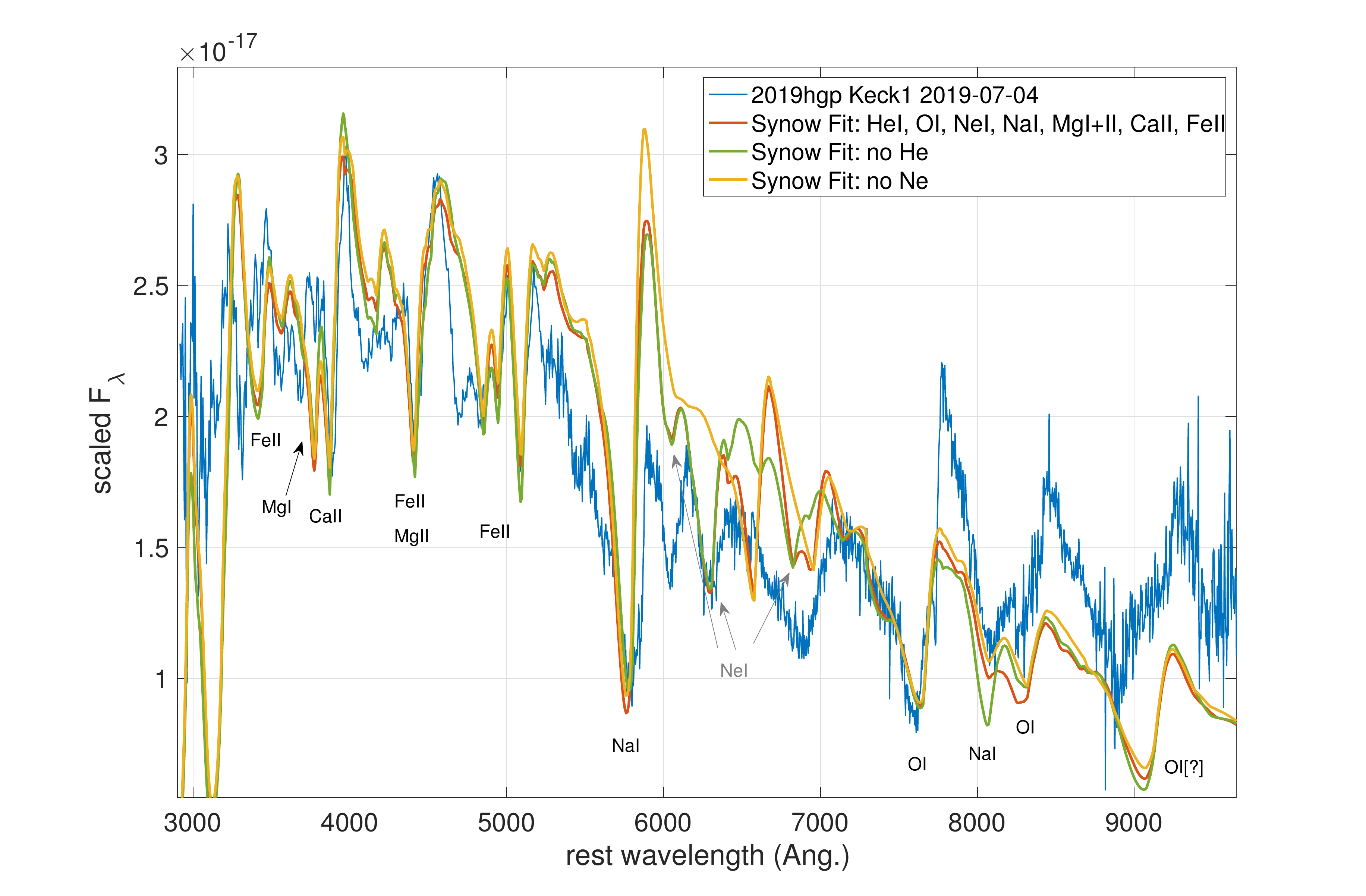}
\vspace*{-0cm}
%\caption{
~\\
{\bf Extended Data Figure 11~~~~}
A comparison of our +27.4\,d Keck spectrum of SN 2019hgp to SYNOW models. The spectrum can be well represented by a combination of common elements seen in supernovae (oxygen, sodium, magnesium, calcium and iron); the addition of neon, which is unique to this object, seems to improve the fit significantly around $6200-7000$\AA\ (yellow).
We compare models without (green) and with (red) He I; we find that the contribution of helium compromises the fit around $6000-7000$\AA, due to the expected but unobserved contribution of the P Cygni profile of He I $\lambda6678$\AA. Perhaps this could be reconciled by more sophisticated modelling, though we note that recent analysis\cite{Karamehmetoglu2019} suggests that the emission component from this particular transition grows stronger with time in spectra of He-rich SNe Ibn. 
%\label{fig:Synow}
%}
\end{figure}

\clearpage

%%%%%%%%%%%%%%%%%%%%%% SI %%%%%%%%%%%%%%%%%

 \item[Supplementary Information]
 \clearpage

\begin{SIfigure}
\centering
\vspace*{-1cm}
\hspace*{-0.5cm}\includegraphics[width=17cm]{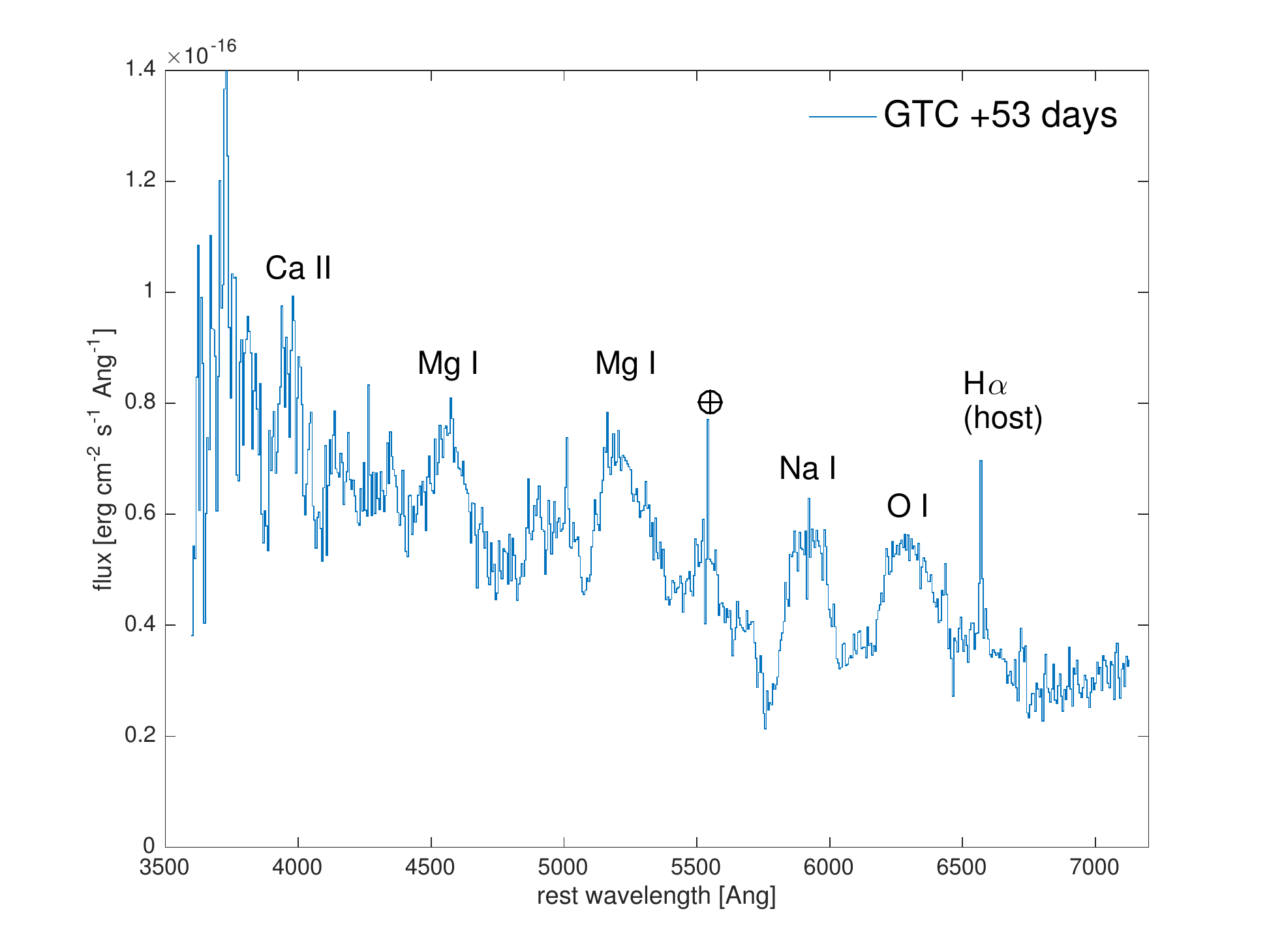}
\vspace*{-8cm}
\caption{
A nebular spectrum of SN 2019hgp obtained 52.8\,d after explosion. Common emission features are marked. Weak absorption from Na I D and and Mg I] $\lambda4571$\AA\ may suggest that the spectrum is not fully nebular at this time.
\label{fig:neb}}
\end{SIfigure}

\clearpage

\begin{SIfigure}
\centering
\vspace*{-1cm}
\hspace*{-0.5cm}\includegraphics[width=17cm]{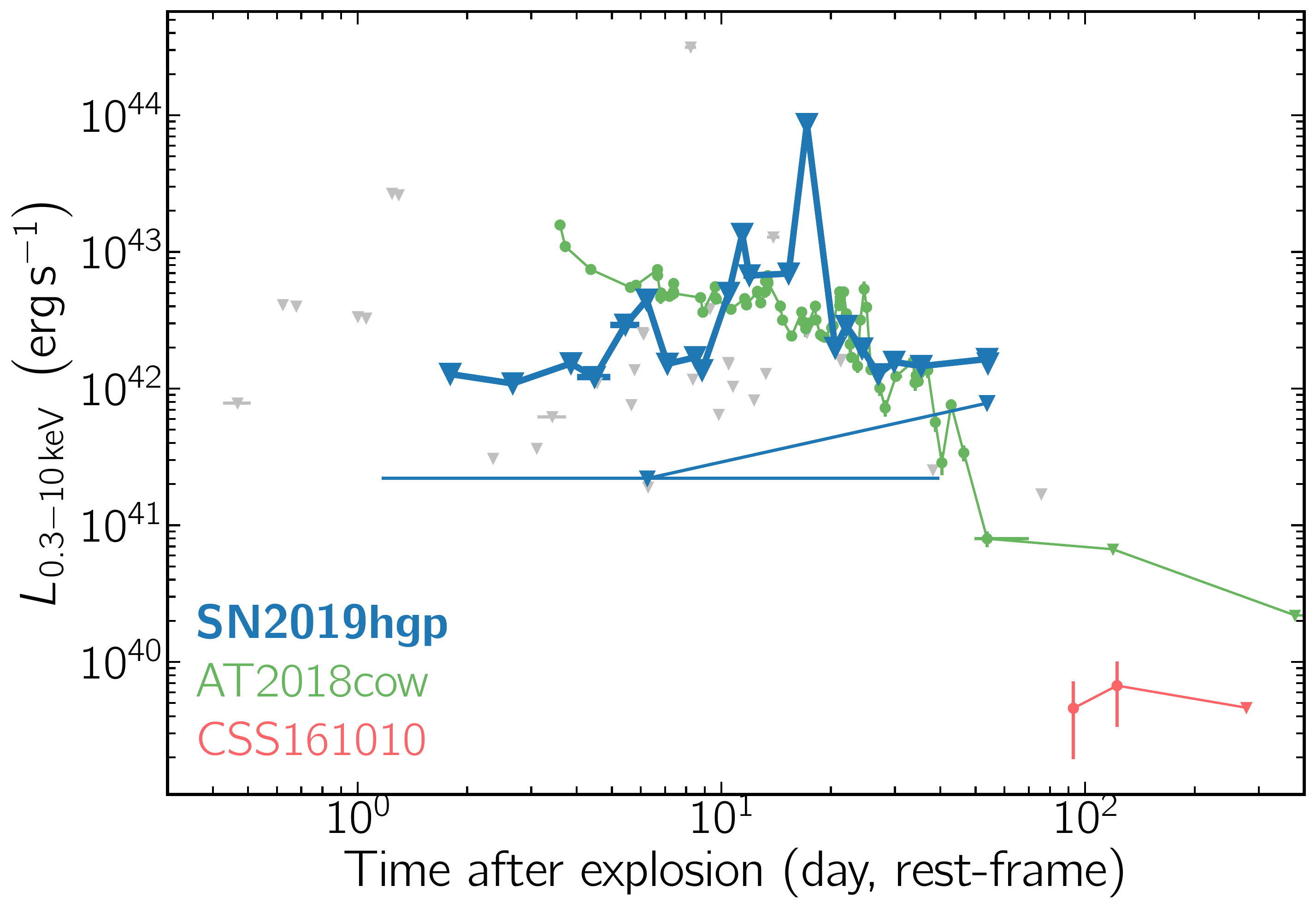}
\vspace*{-2cm}
\caption{Comparison of the X-ray luminosity of RETs. Only two events (AT2018cow and CSS161010; filled circles) are detected. Upper limits (grey triangles - sample; blue triangles - SN 2019hgp) indicate we should have detected SN 2019hgp in X-ray if it had a similar X-ray luminosity to that of AT2018cow, but not if it was similar to CSS161010. For SN 2019hgp, we present orbit stack (connected with a heavy line) as well as dynamically binned limits (thinner line). This comparison motivates more sensitive studies of SNe Ibn and Icn, extending to beyond 100\,days. Standard 1$\sigma$ error bars marked.  
\label{fig:Xsample}}
\end{SIfigure}

\clearpage

\begin{SIfigure}
\centering
\vspace*{-1cm}
\hspace*{-0.5cm}\includegraphics[width=14cm]{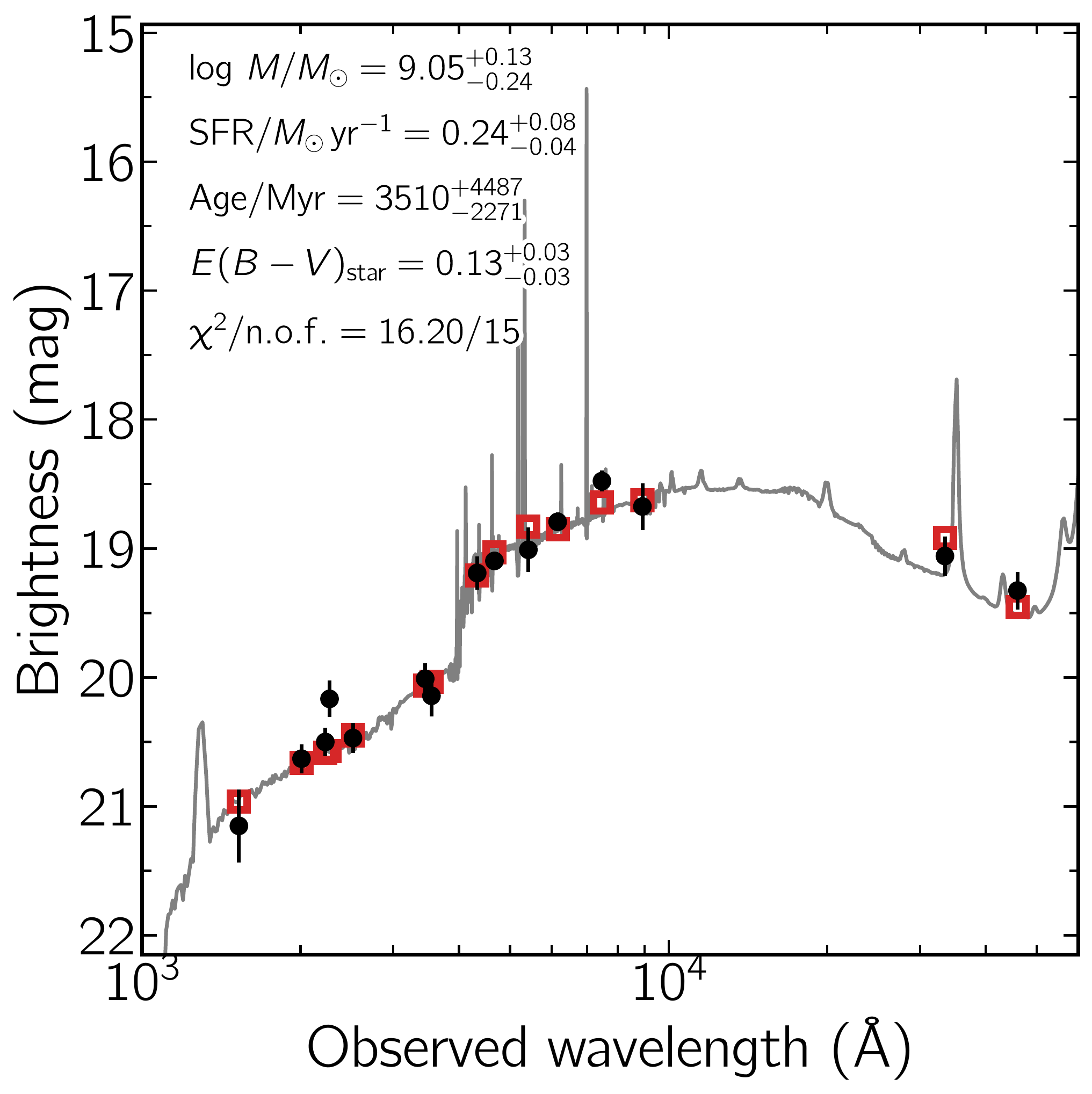}
\vspace*{-2cm}
\caption{
Spectral energy distribution (SED) of the host galaxy of SN 2019hgp from 1000 to 60,000\AA\ (black data points). The solid line displays the best-fitting SED model. The red squares represent the model-predicted magnitudes. The fitting parameters are shown in the upper-left corner. The abbreviation "n.o.f." stands for numbers of filters. Standard 1$\sigma$ error bars marked.
\label{fig:gal_sed}}
\end{SIfigure}

\clearpage

\begin{SIfigure}
\centering
\vspace*{-1cm}
\hspace*{-0.5cm}\includegraphics[width=14cm]{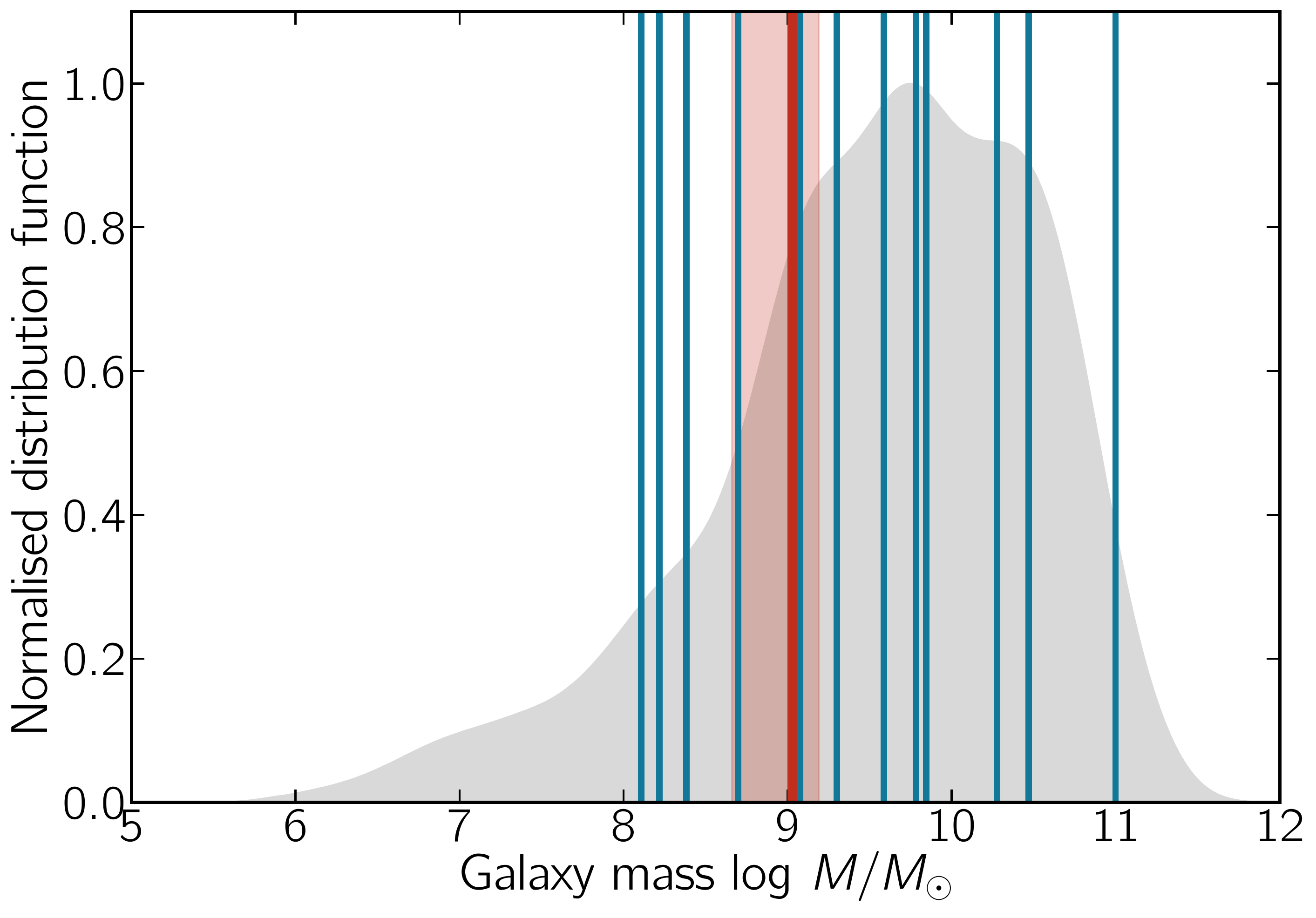}
\vspace*{-2cm}
\caption{
The host-galaxy mass of SN 2019hgp (red line, uncertainty marked with transparent rectangle) and Type II SNe with flash features (blue lines) from the PTF survey in the context of host galaxies of all Type II SNe from the PTF survey (grey distribution). SNe with flash features span a wide range of galaxies from $10^8$\,M$_\odot$ to $10^{11}$\,M$_\odot$. The host of SN 2019hgp does not stand out among the hosts of flash objects.
\label{fig:flash-hosts}}
\end{SIfigure}

\clearpage

\begin{SIfigure}
\centering
\vspace*{-1cm}
\hspace*{-0.5cm}\includegraphics[width=17cm]{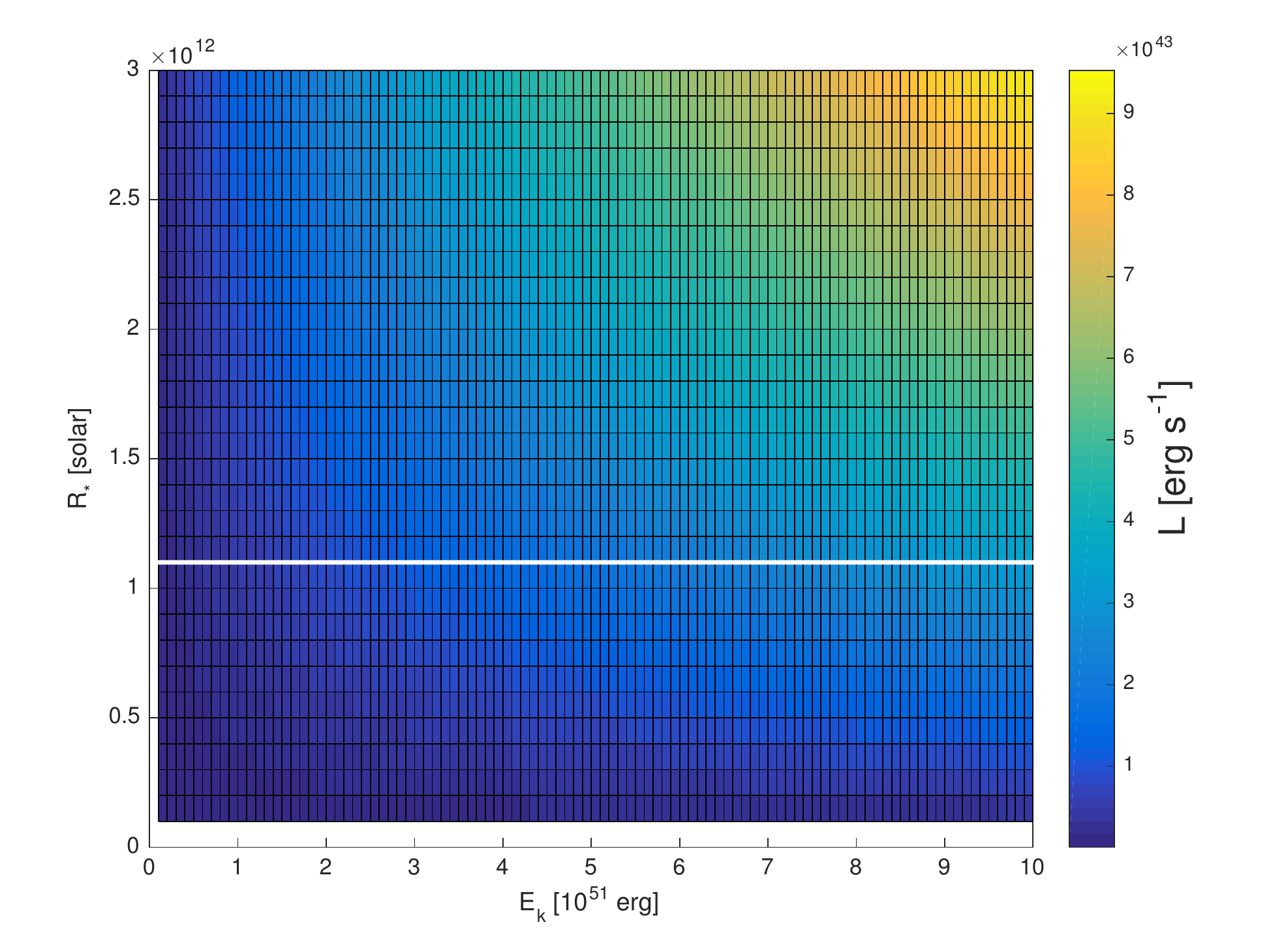}
\vspace*{-8cm}
\caption{
The peak bolometric luminosity of the shock-cooling emission (color-coded) from explosions with varying progenitor radii and explosion energies, calculated using theoretical formulae\cite{Rabinak2011}. The peak luminosity we measure (L$=3.44\times10^{43}$\,erg\,s$^{-1}$; always above white horizontal line) requires supergiant progenitors with R$_{*}>10^{12}$\,cm.  
\label{fig:RW}}
\end{SIfigure}

 %%%%%%%%%%%%%%%%%%%%%%%%%%%%%%%%%%%%%%%%%%%%%%%%%%%%%%%%%%%%%%%%%%%%%%%%%
%								TABLES
%%%%%%%%%%%%%%%%%%%%%%%%%%%%%%%%%%%%%%%%%%%%%%%%%%%%%%%%%%%%%%%%%%%%%%%%%

\clearpage

\begin{deluxetable}{lcccc}
\tablewidth{0pt}
\tabletypesize{\footnotesize}
\tablecaption{SN 2019hgp Photometry}

\tablehead{
\colhead{MJD} & 
\colhead{Rest-frame Phase } & 
\colhead{AB MAG} & 
\colhead{Instrument} & 
\colhead{Filter} \\
\colhead{(days)} &
\colhead{(days)} & 
\colhead{} & 
\colhead{} & 
\colhead{}
}

\startdata
58638.23 & -2.69 & $>20.81$ & P48 & r \\
58638.26 & -2.67 & $>20.78$ & P48 & r \\
58638.26 & -2.67 & $>20.66$ & P48 & r \\
58638.27 & -2.66 & $>20.81$ & P48 & r \\
58638.33 & -2.61 & $>20.81$ & P48 & g \\
58638.33 & -2.6 & $>20.88$ & P48 & g \\
58638.34 & -2.59 & $>20.67$ & P48 & g \\
58638.36 & -2.57 & $>20.72$ & P48 & g \\
58639.21 & -1.77 & $>20.91$ & P48 & g \\
58639.25 & -1.74 & $>20.88$ & P48 & g \\
58639.27 & -1.72 & $>20.85$ & P48 & g \\
58639.28 & -1.71 & $>20.74$ & P48 & r \\
58639.31 & -1.68 & $>20.74$ & P48 & r \\
58639.32 & -1.67 & $>20.73$ & P48 & r \\
58640.24 & -0.8 & $>20.71$ & P48 & r \\
58640.27 & -0.78 & $>20.6$ & P48 & r \\
58640.29 & -0.76 & $>20.82$ & P48 & r \\
58640.32 & -0.73 & $>20.82$ & P48 & g \\
58640.34 & -0.71 & $>20.85$ & P48 & g \\
58640.36 & -0.69 & $>20.86$ & P48 & g \\
58641.31 & 0.19 & $21.95\pm 0.19$ & P48 & r \\
58642.24 & 1.07 & $20.16\pm 0.16$ & P48 & r \\
58642.26 & 1.09 & $20.43\pm 0.21$ & P48 & r \\
58642.29 & 1.12 & $20.26\pm 0.19$ & P48 & r \\
58642.3 & 1.13 & $19.66\pm 0.11$ & P48 & g \\
58642.39 & 1.21 & $19.52\pm 0.12$ & P48 & g \\
58642.43 & 1.25 & $19.91\pm 0.07$ & P60 & r \\
58642.62 & 1.43 & $18.62\pm 0.08$ & SWIFT & UVW2 \\
58642.62 & 1.43 & $18.87\pm 0.09$ & SWIFT & UVW1 \\
58642.62 & 1.43 & $19.14\pm 0.13$ & SWIFT & U \\
58642.62 & 1.43 & $19.42\pm 0.19$ & SWIFT & B \\
58642.63 & 1.43 & $18.78\pm 0.06$ & SWIFT & UVM2 \\
58642.97 & 1.76 & $18.81\pm 0.05$ & LT & u \\
58643.01 & 1.79 & $19.23\pm 0.02$ & LT & g \\
58643.01 & 1.79 & $20.03\pm 0.06$ & LT & i \\
58643.01 & 1.79 & $20.19\pm 0.08$ & LT & z \\
58643.01 & 1.79 & $19.66\pm 0.02$ & LT & r \\
58643.09 & 1.87 & $18.76\pm 0.07$ & LT & u \\
58643.17 & 1.95 & $19.57\pm 0.05$ & P60 & r \\
58643.18 & 1.95 & $19.89\pm 0.05$ & P60 & i \\
58643.20 & 1.98 & $19.14\pm 0.11$ & P48 & g \\
58643.27 & 2.04 & $19.53\pm 0.12$ & P48 & r \\
58643.29 & 2.06 & $19.55\pm 0.14$ & P48 & r \\
58643.33 & 2.09 & $19.05\pm 0.08$ & P48 & g \\
58643.35 & 2.11 & $19.05\pm 0.08$ & P48 & g \\
58643.46 & 2.22 & $19.38\pm 0.07$ & P60 & r \\
58643.55 & 2.31 & $18.52\pm 0.06$ & SWIFT & UVM2 \\
58643.55 & 2.31 & $19.25\pm 0.31$ & SWIFT & V \\
58643.55 & 2.3 & $18.78\pm 0.09$ & SWIFT & UVW1 \\
58643.55 & 2.3 & $18.83\pm 0.13$ & SWIFT & B \\
58643.55 & 2.3 & $18.63\pm 0.1$ & SWIFT & U \\
58643.55 & 2.3 & $18.47\pm 0.08$ & SWIFT & UVW2 \\
58643.93 & 2.66 & $18.89\pm 0.02$ & LT & g \\
58643.93 & 2.66 & $19.27\pm 0.04$ & LT & r \\
58643.93 & 2.66 & $19.7\pm 0.03$ & LT & i \\
58643.93 & 2.66 & $19.88\pm 0.05$ & LT & z \\
58643.93 & 2.66 & $18.63\pm 0.07$ & LT & u \\
58644.14 & 2.86 & $18.66\pm 0.08$ & LT & u \\
58644.19 & 2.9 & $18.79\pm 0.09$ & P48 & g \\
58644.21 & 2.92 & $18.83\pm 0.11$ & P48 & g \\
58644.21 & 2.92 & $18.85\pm 0.09$ & P48 & g \\
58644.22 & 2.93 & $18.8\pm 0.08$ & P48 & g \\
58644.27 & 2.98 & $19.19\pm 0.1$ & P48 & r \\
58644.27 & 2.98 & $19.18\pm 0.1$ & P48 & r \\
58644.29 & 3.0 & $19.25\pm 0.1$ & P48 & r \\
58644.31 & 3.02 & $19.15\pm 0.11$ & P48 & r \\
58644.81 & 3.49 & $18.72\pm 0.1$ & SWIFT & B \\
58644.81 & 3.49 & $18.53\pm 0.07$ & SWIFT & UVW1 \\
58644.81 & 3.49 & $18.45\pm 0.08$ & SWIFT & U \\
58644.82 & 3.49 & $18.43\pm 0.07$ & SWIFT & UVW2 \\
58644.85 & 3.52 & $18.34\pm 0.09$ & SWIFT & UVM2 \\
58644.88 & 3.55 & $19.06\pm 0.31$ & SWIFT & V \\
58644.9 & 3.58 & $18.7\pm 0.02$ & LT & g \\
58644.91 & 3.58 & $19.05\pm 0.03$ & LT & r \\
58644.91 & 3.58 & $19.31\pm 0.05$ & LT & i \\
58644.91 & 3.58 & $19.56\pm 0.1$ & LT & z \\
58644.91 & 3.58 & $18.41\pm 0.07$ & LT & u \\
58645.22 & 3.87 & $18.65\pm 0.08$ & P48 & g \\
58645.24 & 3.89 & $18.62\pm 0.1$ & P48 & g \\
58645.25 & 3.9 & $18.97\pm 0.04$ & P60 & r \\
58645.25 & 3.9 & $18.76\pm 0.04$ & P60 & g \\
58645.25 & 3.9 & $19.2\pm 0.05$ & P60 & i \\
58645.26 & 3.91 & $19.34\pm 0.14$ & P48 & i \\
58645.28 & 3.93 & $19.0\pm 0.1$ & P48 & r \\
58645.3 & 3.94 & $19.02\pm 0.11$ & P48 & r \\
58645.32 & 3.96 & $19.03\pm 0.1$ & P48 & r \\
58645.47 & 4.11 & $18.4\pm 0.07$ & SWIFT & UVW1 \\
58645.47 & 4.11 & $18.55\pm 0.09$ & SWIFT & U \\
58645.48 & 4.11 & $18.62\pm 0.11$ & SWIFT & B \\
58645.48 & 4.11 & $18.45\pm 0.07$ & SWIFT & UVW2 \\
58645.48 & 4.11 & $18.84\pm 0.23$ & SWIFT & V \\
58645.48 & 4.12 & $18.39\pm 0.06$ & SWIFT & UVM2 \\
58645.92 & 4.53 & $19.57\pm 0.11$ & LT & z \\
58645.92 & 4.53 & $18.4\pm 0.08$ & LT & u \\
58645.92 & 4.53 & $19.21\pm 0.03$ & LT & i \\
58645.92 & 4.53 & $18.61\pm 0.01$ & LT & g \\
58645.92 & 4.53 & $18.91\pm 0.02$ & LT & r \\
58646.2 & 4.8 & $18.64\pm 0.09$ & P48 & g \\
58646.24 & 4.83 & $18.91\pm 0.1$ & P48 & r \\
58646.25 & 4.84 & $19.06\pm 0.13$ & P48 & r \\
58646.29 & 4.88 & $18.95\pm 0.1$ & P48 & r \\
58646.32 & 4.91 & $18.71\pm 0.08$ & P48 & g \\
58646.5 & 5.08 & $18.7\pm 0.08$ & SWIFT & UVW2 \\
58646.5 & 5.08 & $18.6\pm 0.19$ & SWIFT & V \\
58646.5 & 5.08 & $18.43\pm 0.09$ & SWIFT & U \\
58646.5 & 5.08 & $18.64\pm 0.12$ & SWIFT & B \\
58646.5 & 5.08 & $18.65\pm 0.07$ & SWIFT & UVW1 \\
58646.51 & 5.08 & $18.56\pm 0.06$ & SWIFT & UVM2 \\
58646.96 & 5.51 & $18.81\pm 0.03$ & LT & r \\
58646.96 & 5.51 & $19.15\pm 0.03$ & LT & i \\
58646.96 & 5.51 & $18.47\pm 0.09$ & LT & u \\
58646.96 & 5.51 & $18.67\pm 0.02$ & LT & g \\
58646.97 & 5.51 & $19.41\pm 0.07$ & LT & z \\
58647.22 & 5.75 & $18.67\pm 0.1$ & P48 & g \\
58647.22 & 5.75 & $18.64\pm 0.1$ & P48 & g \\
58647.25 & 5.78 & $18.79\pm 0.1$ & P48 & r \\
58647.25 & 5.78 & $18.81\pm 0.11$ & P48 & r \\
58647.27 & 5.79 & $18.79\pm 0.09$ & P48 & r \\
58647.27 & 5.8 & $18.95\pm 0.09$ & P48 & r \\
58647.28 & 5.81 & $18.62\pm 0.09$ & P60 & r \\
58647.29 & 5.82 & $18.83\pm 0.09$ & P48 & r \\
58647.3 & 5.83 & $18.67\pm 0.2$ & SWIFT & B \\
58647.3 & 5.83 & $18.98\pm 0.11$ & SWIFT & UVW2 \\
58647.3 & 5.83 & $18.58\pm 0.33$ & SWIFT & V \\
58647.3 & 5.83 & $18.88\pm 0.09$ & SWIFT & UVM2 \\
58647.34 & 5.86 & $18.72\pm 0.08$ & SWIFT & UVW1 \\
58647.34 & 5.86 & $18.66\pm 0.12$ & SWIFT & U \\
58647.34 & 5.87 & $18.73\pm 0.08$ & P48 & g \\
58647.93 & 6.42 & $19.21\pm 0.04$ & LT & z \\
58647.93 & 6.42 & $18.56\pm 0.09$ & LT & u \\
58647.93 & 6.42 & $18.72\pm 0.03$ & LT & g \\
58647.93 & 6.42 & $18.82\pm 0.04$ & LT & r \\
58647.93 & 6.42 & $19.0\pm 0.03$ & LT & i \\
58648.22 & 6.69 & $18.82\pm 0.14$ & P48 & r \\
58648.24 & 6.71 & $18.89\pm 0.12$ & P48 & r \\
58648.26 & 6.73 & $18.86\pm 0.09$ & SWIFT & UVW1 \\
58648.27 & 6.73 & $18.62\pm 0.1$ & SWIFT & U \\
58648.27 & 6.73 & $18.8\pm 0.14$ & SWIFT & B \\
58648.27 & 6.74 & $19.26\pm 0.09$ & SWIFT & UVW2 \\
58648.27 & 6.74 & $18.48\pm 0.2$ & SWIFT & V \\
58648.27 & 6.74 & $19.05\pm 0.08$ & SWIFT & UVM2 \\
58648.32 & 6.79 & $18.83\pm 0.11$ & P48 & g \\
58648.34 & 6.8 & $18.67\pm 0.1$ & P48 & g \\
58648.36 & 6.83 & $18.77\pm 0.12$ & P48 & g \\
58648.37 & 6.83 & $18.65\pm 0.07$ & P60 & r \\
58648.92 & 7.35 & $18.57\pm 0.08$ & LT & u \\
58648.93 & 7.35 & $18.85\pm 0.03$ & LT & r \\
58648.93 & 7.35 & $19.16\pm 0.03$ & LT & z \\
58648.93 & 7.35 & $18.74\pm 0.02$ & LT & g \\
58648.93 & 7.35 & $18.98\pm 0.04$ & LT & i \\
58649.2 & 7.62 & $18.73\pm 0.11$ & P48 & g \\
58649.25 & 7.66 & $18.86\pm 0.11$ & P48 & r \\
58649.26 & 7.67 & $18.78\pm 0.12$ & P48 & r \\
58649.29 & 7.7 & $18.79\pm 0.11$ & P48 & r \\
58649.31 & 7.71 & $18.64\pm 0.04$ & P60 & r \\
58649.31 & 7.72 & $18.78\pm 0.05$ & P60 & i \\
58649.34 & 7.75 & $18.69\pm 0.1$ & P48 & g \\
58649.72 & 8.1 & $18.93\pm 0.09$ & SWIFT & UVW1 \\
58649.72 & 8.11 & $18.51\pm 0.1$ & SWIFT & U \\
58649.73 & 8.11 & $19.59\pm 0.11$ & SWIFT & UVW2 \\
58649.73 & 8.11 & $19.37\pm 0.09$ & SWIFT & UVM2 \\
58649.73 & 8.11 & $18.83\pm 0.15$ & SWIFT & B \\
58649.94 & 8.31 & $18.67\pm 0.04$ & LT & g \\
58649.94 & 8.31 & $18.79\pm 0.04$ & LT & r \\
58649.94 & 8.31 & $19.01\pm 0.06$ & LT & i \\
58649.94 & 8.31 & $18.65\pm 0.1$ & LT & u \\
58649.95 & 8.31 & $19.29\pm 0.07$ & LT & z \\
58650.04 & 8.4 & $18.59\pm 0.08$ & LT & u \\
58650.04 & 8.4 & $19.12\pm 0.06$ & LT & z \\
58650.04 & 8.4 & $18.75\pm 0.02$ & LT & r \\
58650.04 & 8.4 & $18.99\pm 0.02$ & LT & i \\
58650.04 & 8.4 & $18.76\pm 0.03$ & LT & g \\
58650.12 & 8.48 & $19.1\pm 0.1$ & SWIFT & UVW1 \\
58650.12 & 8.48 & $18.58\pm 0.1$ & SWIFT & U \\
58650.12 & 8.48 & $18.64\pm 0.13$ & SWIFT & B \\
58650.13 & 8.48 & $19.63\pm 0.1$ & SWIFT & UVW2 \\
58650.13 & 8.48 & $18.47\pm 0.21$ & SWIFT & V \\
58650.13 & 8.49 & $19.37\pm 0.09$ & SWIFT & UVM2 \\
58650.2 & 8.55 & $18.63\pm 0.09$ & P48 & g \\
58650.2 & 8.55 & $18.72\pm 0.11$ & P48 & g \\
58650.22 & 8.57 & $18.64\pm 0.1$ & P48 & g \\
58650.24 & 8.59 & $18.79\pm 0.1$ & P48 & r \\
58650.24 & 8.59 & $18.72\pm 0.11$ & P48 & r \\
58650.27 & 8.61 & $18.72\pm 0.11$ & P48 & r \\
58650.28 & 8.62 & $18.83\pm 0.11$ & P48 & r \\
58650.96 & 9.26 & $18.81\pm 0.02$ & LT & r \\
58650.96 & 9.26 & $18.99\pm 0.02$ & LT & i \\
58650.96 & 9.27 & $19.19\pm 0.06$ & LT & z \\
58650.96 & 9.27 & $18.73\pm 0.09$ & LT & u \\
58651.19 & 9.48 & $18.86\pm 0.13$ & P48 & g \\
58651.21 & 9.5 & $18.9\pm 0.15$ & P48 & g \\
58651.22 & 9.51 & $18.78\pm 0.11$ & P48 & g \\
58651.23 & 9.52 & $18.82\pm 0.04$ & P60 & r \\
58651.24 & 9.53 & $18.77\pm 0.11$ & P48 & r \\
58651.24 & 9.52 & $18.89\pm 0.04$ & P60 & i \\
58651.28 & 9.56 & $18.8\pm 0.11$ & P48 & r \\
58651.29 & 9.58 & $18.77\pm 0.14$ & P48 & r \\
58651.85 & 10.11 & $18.88\pm 0.12$ & SWIFT & U \\
58651.85 & 10.1 & $19.51\pm 0.12$ & SWIFT & UVW1 \\
58651.85 & 10.11 & $19.0\pm 0.16$ & SWIFT & B \\
58651.86 & 10.11 & $20.12\pm 0.13$ & SWIFT & UVW2 \\
58651.86 & 10.11 & $18.75\pm 0.25$ & SWIFT & V \\
58651.86 & 10.11 & $19.67\pm 0.1$ & SWIFT & UVM2 \\
58651.92 & 10.17 & $18.82\pm 0.01$ & LT & g \\
58651.92 & 10.17 & $18.83\pm 0.01$ & LT & r \\
58651.92 & 10.17 & $18.98\pm 0.03$ & LT & i \\
58651.92 & 10.17 & $19.16\pm 0.09$ & LT & z \\
58651.93 & 10.17 & $18.68\pm 0.09$ & LT & u \\
58652.18 & 10.41 & $18.78\pm 0.12$ & P48 & g \\
58652.2 & 10.44 & $18.91\pm 0.15$ & P48 & g \\
58652.22 & 10.45 & $18.8\pm 0.12$ & P48 & g \\
58652.29 & 10.52 & $18.79\pm 0.09$ & P48 & r \\
58652.3 & 10.52 & $18.85\pm 0.12$ & P48 & r \\
58652.32 & 10.55 & $18.91\pm 0.12$ & P48 & r \\
58652.85 & 11.05 & $19.93\pm 0.14$ & SWIFT & UVM2 \\
58652.85 & 11.04 & $19.16\pm 0.46$ & SWIFT & V \\
58652.85 & 11.04 & $18.65\pm 0.18$ & SWIFT & B \\
58652.85 & 11.04 & $18.74\pm 0.15$ & SWIFT & U \\
58652.85 & 11.04 & $19.62\pm 0.16$ & SWIFT & UVW1 \\
58652.85 & 11.04 & $19.96\pm 0.15$ & SWIFT & UVW2 \\
58652.92 & 11.11 & $18.86\pm 0.01$ & LT & g \\
58652.92 & 11.11 & $18.85\pm 0.02$ & LT & r \\
58652.92 & 11.11 & $19.0\pm 0.02$ & LT & i \\
58652.93 & 11.11 & $19.1\pm 0.04$ & LT & z \\
58652.93 & 11.11 & $19.23\pm 0.04$ & LT & z \\
58652.93 & 11.12 & $18.84\pm 0.07$ & LT & u \\
58653.2 & 11.37 & $18.85\pm 0.1$ & P48 & r \\
58653.22 & 11.39 & $18.86\pm 0.1$ & P48 & r \\
58653.25 & 11.42 & $18.9\pm 0.11$ & P48 & i \\
58653.27 & 11.43 & $18.95\pm 0.12$ & P48 & g \\
58653.29 & 11.46 & $18.91\pm 0.15$ & P48 & g \\
58653.29 & 11.46 & $18.9\pm 0.14$ & P48 & g \\
58653.3 & 11.47 & $18.79\pm 0.11$ & P48 & g \\
58653.41 & 11.57 & $19.96\pm 0.12$ & SWIFT & UVM2 \\
58653.41 & 11.57 & $19.2\pm 0.37$ & SWIFT & V \\
58653.41 & 11.57 & $18.81\pm 0.15$ & SWIFT & B \\
58653.41 & 11.57 & $18.92\pm 0.13$ & SWIFT & U \\
58653.41 & 11.57 & $20.41\pm 0.15$ & SWIFT & UVW2 \\
58653.41 & 11.57 & $19.58\pm 0.13$ & SWIFT & UVW1 \\
58653.93 & 12.05 & $18.9\pm 0.02$ & LT & g \\
58653.93 & 12.06 & $18.98\pm 0.01$ & LT & r \\
58653.93 & 12.06 & $19.07\pm 0.01$ & LT & i \\
58653.93 & 12.06 & $19.18\pm 0.04$ & LT & z \\
58653.93 & 12.06 & $19.07\pm 0.06$ & LT & u \\
58654.21 & 12.32 & $18.98\pm 0.11$ & P48 & r \\
58654.22 & 12.33 & $18.89\pm 0.11$ & P48 & r \\
58654.25 & 12.36 & $18.98\pm 0.11$ & P48 & g \\
58654.26 & 12.37 & $18.93\pm 0.11$ & P48 & g \\
58654.29 & 12.4 & $18.94\pm 0.14$ & P48 & g \\
58655.19 & 13.24 & $19.03\pm 0.11$ & P48 & r \\
58655.2 & 13.25 & $19.05\pm 0.1$ & P48 & r \\
58655.22 & 13.27 & $19.05\pm 0.11$ & P48 & r \\
58655.24 & 13.29 & $19.09\pm 0.11$ & P48 & g \\
58655.26 & 13.3 & $18.97\pm 0.14$ & P48 & g \\
58655.28 & 13.33 & $19.0\pm 0.13$ & P48 & g \\
58656.92 & 14.87 & $19.17\pm 0.02$ & LT & r \\
58656.92 & 14.87 & $19.2\pm 0.02$ & LT & g \\
58656.93 & 14.87 & $19.18\pm 0.02$ & LT & r \\
58656.93 & 14.87 & $19.2\pm 0.02$ & LT & i \\
58656.93 & 14.88 & $19.2\pm 0.03$ & LT & z \\
58656.93 & 14.88 & $19.41\pm 0.07$ & LT & u \\
58656.99 & 14.94 & $20.45\pm 0.4$ & SWIFT & UVW1 \\
58656.99 & 14.94 & $19.35\pm 0.31$ & SWIFT & U \\
58656.99 & 14.94 & $18.79\pm 0.3$ & SWIFT & B \\
58657.0 & 14.94 & $20.61\pm 0.29$ & SWIFT & UVW2 \\
58657.0 & 14.94 & $20.76\pm 0.3$ & SWIFT & UVM2 \\
58657.18 & 15.11 & $19.26\pm 0.15$ & P48 & i \\
58657.21 & 15.14 & $19.21\pm 0.15$ & P48 & g \\
58657.23 & 15.15 & $19.25\pm 0.13$ & P48 & g \\
58657.23 & 15.16 & $19.3\pm 0.1$ & P48 & g \\
58657.26 & 15.19 & $19.24\pm 0.11$ & P48 & r \\
58657.26 & 15.19 & $19.25\pm 0.13$ & P48 & r \\
58657.35 & 15.27 & $19.21\pm 0.1$ & P48 & r \\
58657.37 & 15.29 & $19.26\pm 0.13$ & P48 & r \\
58657.92 & 15.81 & $19.64\pm 0.11$ & LT & u \\
58657.92 & 15.81 & $19.23\pm 0.07$ & LT & z \\
58657.92 & 15.8 & $19.27\pm 0.02$ & LT & r \\
58657.92 & 15.81 & $19.28\pm 0.03$ & LT & i \\
58657.92 & 15.8 & $19.31\pm 0.02$ & LT & g \\
58658.22 & 16.09 & $19.38\pm 0.11$ & P48 & g \\
58658.24 & 16.11 & $19.37\pm 0.1$ & P48 & r \\
58658.92 & 16.75 & $19.71\pm 0.13$ & LT & u \\
58658.92 & 16.75 & $19.3\pm 0.04$ & LT & z \\
58658.92 & 16.74 & $19.42\pm 0.02$ & LT & g \\
58658.92 & 16.74 & $19.38\pm 0.02$ & LT & r \\
58658.92 & 16.75 & $19.36\pm 0.03$ & LT & i \\
58659.45 & 17.25 & $19.39\pm 0.08$ & P60 & r \\
58659.46 & 17.25 & $19.28\pm 0.05$ & P60 & i \\
58660.18 & 17.93 & $19.46\pm 0.17$ & P48 & g \\
58660.2 & 17.95 & $19.53\pm 0.04$ & P60 & r \\
58660.2 & 17.95 & $19.55\pm 0.11$ & P48 & g \\
58660.21 & 17.96 & $19.62\pm 0.03$ & P60 & g \\
58660.21 & 17.96 & $19.28\pm 0.05$ & P60 & i \\
58660.24 & 17.99 & $19.49\pm 0.1$ & P48 & r \\
58660.24 & 17.99 & $19.53\pm 0.09$ & P48 & r \\
58661.18 & 18.87 & $19.64\pm 0.22$ & P48 & i \\
58661.31 & 18.99 & $19.6\pm 0.14$ & P48 & r \\
58661.91 & 19.56 & $19.61\pm 0.03$ & LT & i \\
58661.91 & 19.56 & $19.85\pm 0.03$ & LT & g \\
58661.91 & 19.56 & $19.7\pm 0.02$ & LT & r \\
58661.92 & 19.56 & $20.44\pm 0.14$ & LT & u \\
58661.92 & 19.56 & $19.55\pm 0.04$ & LT & z \\
58662.22 & 19.85 & $19.88\pm 0.18$ & P48 & g \\
58662.24 & 19.87 & $19.61\pm 0.13$ & P48 & r \\
58662.58 & 20.18 & $21.64\pm 0.29$ & SWIFT & UVW2 \\
58662.58 & 20.19 & $21.7\pm 0.44$ & SWIFT & UVW1 \\
58663.21 & 20.78 & $19.74\pm 0.17$ & P48 & r \\
58663.21 & 20.78 & $19.82\pm 0.17$ & P48 & r \\
58663.22 & 20.79 & $19.96\pm 0.18$ & P48 & g \\
58663.23 & 20.79 & $20.15\pm 0.16$ & P48 & g \\
58663.31 & 20.87 & $19.79\pm 0.16$ & P48 & r \\
58663.9 & 21.42 & $20.04\pm 0.06$ & LT & g \\
58663.9 & 21.42 & $19.95\pm 0.02$ & LT & r \\
58663.9 & 21.43 & $19.84\pm 0.04$ & LT & i \\
58663.9 & 21.43 & $19.64\pm 0.06$ & LT & z \\
58663.9 & 21.43 & $20.43\pm 0.17$ & LT & u \\
58665.2 & 22.65 & $20.14\pm 0.15$ & P48 & g \\
58665.23 & 22.67 & $20.23\pm 0.2$ & P48 & g \\
58665.26 & 22.71 & $20.22\pm 0.19$ & P48 & r \\
58665.89 & 23.3 & $20.17\pm 0.07$ & LT & r \\
58665.89 & 23.3 & $19.79\pm 0.04$ & LT & z \\
58665.89 & 23.29 & $20.34\pm 0.07$ & LT & g \\
58665.89 & 23.3 & $20.07\pm 0.07$ & LT & i \\
58666.22 & 23.61 & $20.27\pm 0.18$ & P48 & g \\
58666.23 & 23.61 & $20.42\pm 0.19$ & P48 & g \\
58666.25 & 23.63 & $20.44\pm 0.23$ & P48 & g \\
58666.26 & 23.65 & $20.25\pm 0.16$ & P48 & r \\
58666.26 & 23.65 & $20.18\pm 0.17$ & P48 & r \\
58667.35 & 24.67 & $20.63\pm 0.28$ & P48 & g \\
58668.18 & 25.45 & $20.46\pm 0.22$ & P48 & r \\
58668.2 & 25.47 & $20.55\pm 0.19$ & P48 & g \\
58668.92 & 26.15 & $20.65\pm 0.06$ & LT & r \\
58668.92 & 26.15 & $20.95\pm 0.07$ & LT & g \\
58668.93 & 26.15 & $20.51\pm 0.07$ & LT & i \\
58668.93 & 26.15 & $20.01\pm 0.1$ & LT & z \\
58670.91 & 28.02 & $20.3\pm 0.09$ & LT & z \\
58670.91 & 28.01 & $21.06\pm 0.07$ & LT & g \\
58670.91 & 28.01 & $20.97\pm 0.07$ & LT & r \\
58670.91 & 28.02 & $20.76\pm 0.07$ & LT & i \\
58674.0 & 30.92 & $21.32\pm 0.13$ & LT & r \\
58674.0 & 30.92 & $21.57\pm 0.19$ & LT & g \\
58674.01 & 30.92 & $21.07\pm 0.12$ & LT & i \\
58674.01 & 30.93 & $20.41\pm 0.16$ & LT & z \\
58675.9 & 32.7 & $21.42\pm 0.28$ & LT & g \\
58675.9 & 32.7 & $21.65\pm 0.19$ & LT & r \\
58675.9 & 32.7 & $21.43\pm 0.24$ & LT & i \\
58675.9 & 32.71 & $20.81\pm 0.15$ & LT & z \\
58681.94 & 38.38 & $21.74\pm 0.47$ & LT & i \\
58681.94 & 38.38 & $21.15\pm 0.33$ & LT & z \\
58684.92 & 41.18 & $22.03\pm 0.41$ & LT & i \\
58684.92 & 41.18 & $22.65\pm 0.38$ & LT & r \\
58684.93 & 41.19 & $21.44\pm 0.29$ & LT & z \\
58687.94 & 44.02 & $21.54\pm 0.22$ & LT & z \\
58693.9 & 49.62 & $23.06\pm 0.04$ & GTC & g \\
58693.9 & 49.62 & $22.72\pm 0.06$ & GTC & i \\
58693.9 & 49.62 & $22.78\pm 0.04$ & GTC & r \\
58694.94 & 50.59 & $22.66\pm 0.06$ & GTC & i \\
58694.94 & 50.6 & $21.99\pm 0.08$ & GTC & z \\
58983.43 & 321.71 & $21.2\pm 0.19$ & P48 & g \\
58984.41 & 322.63 & $21.32\pm 0.11$ & P48 & g \\
\enddata
\label{phot_tab}
\tablecomments{Full table is available as a separate, machine-readable file. A
portion is shown here for clarity.
No extinction correction has been applied. The GTC photometry is not host-subtracted.}

\end{deluxetable}

\clearpage

\begin{deluxetable}{lcccccc}
\tablewidth{0pt}
\tabletypesize{\footnotesize}
\tablecaption{SN 2019hgp Spectroscopy}
\tablehead{
\colhead{Observation date} & 
%\colhead{Time} & 
\colhead{Phase} &
\colhead{Facility} &
\colhead{Exp. time} &
\colhead{Grism/Grating} &
\colhead{Slit} &
\colhead{Range} \\
%\colhead{} & 
\colhead{(UTC)} & 
\colhead{(days)} &
\colhead{} &
\colhead{(s)} &
\colhead{} &
\colhead{(arcsec)} &
\colhead{(\AA)} 
}
\startdata
2019 Jun 08	03:18:05	& 1.0		&	P60/SEDM	&	2250	&	& IFU		&	3776--9200	\\
2019 Jun 08	12:51:13	& 1.4		&	Gemini N./GMOS	&	$2\times900$	&	B600+G5307	&	1	&		3630--6850	\\
2019 Jun 08 12:51:13 (host)	& 1.4		&	Gemini N./GMOS	&	$2\times900$	&	B600+G5307	&	1.0	&	3630--6850	\\
2019 Jun 08	22:03:49	& 1.8		&	LT/SPRAT	&	1200	&		& 1.8		& 	4020--7960	\\
2019 Jun 08	22:12:38	& 1.8		&	P60/SEDM	&	2250	&	&	IFU	&	3776--9200	\\
2019 Jun 09	02:14:49	& 2.0		&	LT/SPRAT	&	1400	&		& 1.8		& 	4020--7960	\\
2019 Jun 09	04:00:39	& 2.1		&	P60/SEDM	&	2250	&	& IFU		&	3776--9200	\\
2019 Jun 09	22:26:20	& 2.8		&	LT/SPRAT	&	1600	&		& 1.8		&	4020--7960	\\
2019 Jun 10	07:26:06	& 3.2		&	Gemini N./GMOS	&	$2\times900$	&	B600+G5307	&	1	&	3630--6850	\\
2019 Jun 10	21:55:03	& 3.8		&	LT/SPRAT	&	600	&		& 1.8		& 	4020--7960	\\
2019 Jun 10	22:05:12	& 3.8		&	LT/SPRAT	&	600	&		& 1.8		&	4020--7960	\\
2019 Jun 11	22:47:15	& 4.8		&	P60/SEDM	&	2250	&	& IFU		&	3776--9200	\\
2019 Jun 11	23:33:50	& 4.9		&	NOT/ALFOSC	&	2700	&	Grism\_\#4	&	1	&	3600--9700	\\
2019 Jun 12	08:03:00	& 5.2		&	HET/LRS2	&	1800	& blue arm	&	IFU	&	3640--6950	\\
2019 Jun 12	22:02:47	& 5.8		&	WHT/ACAM	&	900	&	V400	&	1	&	3750--9200	\\
2019 Jun 12	23:45:21	& 5.9		&	P60/SEDM	&	2250	&	&	IFU	&	3776--9200	\\
%2019 Jun 14	&		&	Lick 3-m/KAST	&		&		&		&	3200 – 9150	\\
2019 Jun 14	01:57:52	& 6.9		&	P60/SEDM	&	2250	&	&	IFU	&	3776--9200	\\
2019 Jun 15	00:39:02	& 7.9		&	P60/SEDM	&	2250	&	&	IFU	&	3776--9200	\\
2019 Jun 15	07:13:00	& 8.2		&	HET/LRS2	&	2000	&	blue arm	& IFU		& 3640--6950	\\
2019 Jun 15	22:02:17	& 8.9		&	LT/SPRAT	&	1800	&		& 1.8		&	4020-–7960	\\
2019 Jun 16	21:53:05	& 9.8		&	P60/SEDM	&	2250	&	& IFU		&	3776-–9200	\\
2019 Jun 16	22:08:36	& 9.8		&	LT/SPRAT	&	2400	&		& 1.8		&  4020-–7960	\\
2019 Jun 17	06:58:00	& 10.2		&	HET/LRS2	&	1800	&	blue arm	& IFU &	3640-–6950	\\
2019 Jun 17	21:44:01	& 10.8		&	NOT/ALFOSC	&	2400	&	Grism\_\#4	&	1	&	3600-–9700	\\
2019 Jun 19	02:31:11	& 12.0		&	P60/SEDM	&	2250	&	& IFU		&	3776-–9200	\\
2019 Jun 19	22:45:24	& 12.8		&	LT/SPRAT	&	1200	&		& 1.8		&	4020-–7960	\\
2019 Jun 19	23:05:33	& 12.9		&	LT/SPRAT	&	1200	&		& 1.8		&	4020-–7960	\\
2019 Jun 22	07:49:13	& 15.2		&	LDT/Deveny/LMI	&	$2\times450$	&	300/4000	&	1.5	&	3550-–7970	\\
2019 Jun 26	09:05:37	& 19.3		&	P200/DBSP	&	1200	&	600/4000 \& 316/7150	&   1.5	&	3600-–10500	\\
2019 Jun 30	09:50:46	& 23.3		&	Keck1/LRIS	&	1800	&		&	1	&	3120-–10230	\\
2019 Jul 01	07:43:10	& 24.2		&	P200/DBSP	&	1500	&	600/4000 \& 316/7150	&	1.5	&	3400-–10000	\\
2019 Jul 04	10:49:25	& 27.4		&	Keck1/LRIS	&	850	&	400/3400, 400/8500	&	1	&	3100-–10300	\\
2019 Jul 29	21:19:22	& 52.8		&	GTC/OSIRIS	&	3x1400	&	R1000B \& R1000R	&	0.8	&	3630-–10200	\\
2019 Jul 30	04:21:00	& 53.1		&	HET/LRS2	&	1800	& red \& blue arm	& IFU	&	4010-–9950	\\
\enddata
\label{spec_tab}
\tablecomments{The phase is calculated with respect to the estimated explosion date on June 7.1 2019 and is here given in the observer frame.}
\end{deluxetable}

\clearpage

\begin{deluxetable}{cccc}
\tablewidth{0pt}
\tabletypesize{\footnotesize}
\tablecaption{Photometry of the host galaxy of SN 2019hgp}
\tablehead{
\colhead{Survey/}& 
\colhead{Instrument}& 
\colhead{Filter}& \colhead{Brightness}\\
\colhead{Telescope}&
\colhead{}&                  
\colhead{}& 
\colhead{(mag)}
}
\startdata
\textit{GALEX}   &           &$ FUV   $&$ 21.31 \pm 0.26 $\\
\textit{Swift}   & UVOT      &$ uvw2  $&$ 20.80 \pm 0.05 $\\
\textit{GALEX}   &           &$ NUV   $&$ 20.34 \pm 0.10 $\\
\textit{Swift}   & UVOT      &$ uvm2  $&$ 20.67 \pm 0.05 $\\
\textit{Swift}   & UVOT      &$ uvw1  $&$ 20.60 \pm 0.06 $\\
\textit{Swift}   & UVOT      &$ u     $&$ 20.11 \pm 0.07 $\\
SDSS             &           &$ u'    $&$ 20.23 \pm 0.15 $\\
\textit{Swift}   & UVOT      &$ b     $&$ 19.26 \pm 0.08 $\\
SDSS             &           &$ g'    $&$ 19.17 \pm 0.03 $\\
\textit{Swift}   & UVOT      &$ v     $&$ 19.07 \pm 0.14 $\\
SDSS             &           &$ r'    $&$ 18.85 \pm 0.05 $\\
SDSS             &           &$ i'    $&$ 18.52 \pm 0.06 $\\
SDSS             &           &$ z'    $&$ 18.70 \pm 0.17 $\\
\textit{WISE}    &           &$ W1    $&$ 19.06 \pm 0.12 $\\
\textit{WISE}    &           &$ W2    $&$ 19.33 \pm 0.11 $\\
\enddata
\label{tab:hostphot}
\tablecomments{All magnitudes are reported in the AB system and not corrected for reddening.}
\end{deluxetable}

\end{addendum}
 
\end{document}